# General Relativity as a fully singular Lagrange system

T. Mei

(Department of Journal, Central China Normal University, Wuhan, Hubei PRO, People's Republic of China
E-Mail: meitao@mail.ccnu.edu.cn meitaowh@public.wh.hb.cn )

**Abstract:** We present some gauge conditions to eliminate all second time derivative terms in the vierbein forms of the ten Einstein equations of general relativity; at the same time, we present the corresponding Lagrangian in which there is not any quadratic term of first time derivative that can leads to those vierbein forms of the Einstein equations without second time derivative term by the corresponding Euler-Lagrange equations. General relativity thus becomes a fully singular Lagrange system.

## 0 Introduction

As well known, if $x^0$ is the time coordinate of a frame of reference, then the ten Einstein equations $R^{\mu\nu} - \frac{1}{2} g^{\mu\nu} R = \frac{8\pi G}{c^4} T^{\mu\nu}$ can be divided into two categories: One is the four equations $R^{0\lambda} - \frac{1}{2} g^{0\lambda} R = \frac{8\pi G}{c^4} T^{0\lambda}$ in which there is not any second time derivative term, another is the six equations $R^{ij} - \frac{1}{2} g^{ij} R = \frac{8\pi G}{c^4} T^{ij}$ including the six second time derivative terms $g_{ij,0,0}$. On the other hand, we have presented a group of gauge conditions to eliminate all second time derivative terms in the vierbein forms of the ten Einstein equations in Ref. [1], however, for which so far I cannot find out the corresponding Lagrangian density. In this paper, we not only present some gauge conditions that can eliminate all second time derivative terms in the vierbein forms of the ten Einstein equations, but also, at the same time, present the corresponding Lagrangian in which there is not any quadratic term of first time derivative that can leads to those vierbein forms of the Einstein equations without second time derivative term by the corresponding Euler-Lagrange equations. General relativity thus becomes a fully singular Lagrange system.

In Sect. 1 of this paper, we present a Lagrangian density of general relativity, which is separated to kinetic energy and potential energy terms naturally; In Sect. 2, we investigate in detail a group of gauge conditions such that general relativity becomes a fully singular Lagrange system; In Sect. 3, we generate the group of gauge conditions discussed in Sect. 2 to more general cases.

All symbols and conventions follow Ref. [1].

## 1 A Lagrangian density of general relativity

As well known, in general relativity, the Einstein-Hilbert action reads

$$S_{\mathrm{EH}}=\frac{c^3}{16\pi G}\int\sqrt{-g}\mathrm{d}^4xR=\frac{c^3}{16\pi G}\int\sqrt{-g}\mathrm{d}^4x\,L_{\mathrm{G}(1)}+\frac{c^3}{16\pi G}\int\mathrm{d}^4x\frac{\partial}{\partial x^\mu}\left[\sqrt{-g}(g^{\rho\sigma}\Gamma^\mu_{\rho\sigma}-g^{\mu\rho}\Gamma^\sigma_{\rho\sigma})\right], \quad (1\text{-}1)$$

$$L_{\mathrm{G}(1)}=g^{\alpha\beta}(\Gamma^\rho_{\alpha\sigma}\Gamma^\sigma_{\beta\rho}-\Gamma^\rho_{\alpha\beta}\Gamma^\sigma_{\rho\sigma}). \quad (1\text{-}2)$$

We introduce $\tilde{g}^{ij}=g^{ij}-\frac{g^{0i}g^{0j}}{g^{00}}$, which satisfies $\tilde{g}^{ik}g_{kj}=\delta^i_j$, $\tilde{g}^{ij}$ is thus the inverse matrix of $g_{ij}$, and, further, the 3-dimentional Christoffel symbol can be written to the form $\tilde{\Gamma}^i_{lm}=\frac{1}{2}\tilde{g}^{ij}\left(g_{jl,m}+g_{jm,l}-g_{lm,j}\right)$. Based on $\tilde{g}^{ij}$ and $\tilde{\Gamma}^i_{lm}$, we define

$$L_{\mathrm{GK}}=\frac{1}{-g^{00}}\left(\tilde{g}^{il}\tilde{g}^{jm}-\tilde{g}^{ij}\tilde{g}^{lm}\right)\Gamma^0_{ij}\Gamma^0_{lm}, \quad (1\text{-}3)$$

$$\begin{aligned}L_{\mathrm{GU}}&=\tilde{g}^{ij}\left(\frac{g^{0\lambda}}{g^{00}}\left(\Gamma^0_{\lambda k}\tilde{\Gamma}^k_{ij}-\Gamma^0_{\lambda j}\tilde{\Gamma}^k_{ik}\right)-\left(\tilde{\Gamma}^l_{im}\tilde{\Gamma}^m_{jl}-\tilde{\Gamma}^l_{ij}\tilde{\Gamma}^m_{lm}\right)\right)\\&=\frac{1}{2}\frac{1}{g^{00}}g^{0\rho}g^{0\sigma}g_{\rho\sigma,k}\left(\tilde{g}^{ij}\tilde{\Gamma}^k_{ij}-\tilde{g}^{ik}\tilde{\Gamma}^j_{ij}\right)-\tilde{g}^{ij}\left(\tilde{\Gamma}^l_{im}\tilde{\Gamma}^m_{jl}-\tilde{\Gamma}^l_{ij}\tilde{\Gamma}^m_{lm}\right) \quad (1\text{-}4)\\&=-\frac{1}{2}\frac{g^{00}{}_{,k}}{g^{00}}\left(\tilde{g}^{ij}\tilde{\Gamma}^k_{ij}-\tilde{g}^{ik}\tilde{\Gamma}^j_{ij}\right)-\tilde{g}^{ij}\left(\tilde{\Gamma}^l_{im}\tilde{\Gamma}^m_{jl}-\tilde{\Gamma}^l_{ij}\tilde{\Gamma}^m_{lm}\right);\end{aligned}$$

It can be proved that the remainder of $\sqrt{-g}L_{\mathrm{G}(1)}$ and $\sqrt{-g}\left(L_{\mathrm{GK}}-L_{\mathrm{GU}}\right)$ is a total derivative:

$$\sqrt{-g}L_{\mathrm{G}(1)}=\sqrt{-g}\left(L_{\mathrm{GK}}-L_{\mathrm{GU}}\right)+\frac{\partial}{\partial x^\mu}\left[\frac{\sqrt{-g}}{g^{00}}\left(g^{0\nu}\frac{\partial g^{0\mu}}{\partial x^\nu}-g^{0\mu}\frac{\partial g^{0\nu}}{\partial x^\nu}\right)\right], \quad (1\text{-}5)$$

we therefore can use

$$L_{\mathrm{G}(2)}=L_{\mathrm{GK}}-L_{\mathrm{GU}} \quad (1\text{-}6)$$

as a Lagrangian density of general relativity.

As well known, the coordinate variables $\left(x^0,x^1,x^2,x^3\right)$ in general relativity are four parameters. In the following discussion, we shall assume that the decomposition of time and space has been finished by the ADM decomposition[2] and by $x^0$ we denote time coordinate.

If $x^0$ is the time coordinate of a frame of reference, from (1-4) we see that there is not any time derivative term in $L_{\mathrm{GU}}$, and all time derivative terms appear in $L_{\mathrm{GK}}$; hence, $L_{\mathrm{GK}}$ and $L_{\mathrm{GU}}$ can be regarded as kinetic energy and potential energy terms, respectively. Especially, for the metric of an accelerated, rotating frame of reference [3, 4, 5]

$$g_{00}=-\left(\left(1+\frac{\boldsymbol{W}(t)\cdot\boldsymbol{r}}{c^2}\right)^2-\left(\frac{\boldsymbol{\varpi}(t)\times\boldsymbol{r}}{c}\right)^2\right),\quad g_{0i}=g_{i0}=\left(\frac{\boldsymbol{\varpi}(t)\times\boldsymbol{r}}{c}\right)_i,\quad g_{ij}=\delta_{ij}, \quad (1\text{-}7)$$

we have $L_{\rm GK} = L_{\rm GU} = 0$ since $\Gamma^0_{ij} = 0$ and $\widetilde{\Gamma}^i_{lm} = 0$, this is a due result, because there is not real gravitational field for an accelerated, rotating frame of reference.

## 2 General relativity as a fully singular Lagrange system under a group of gauge conditions

### 2.1 The vierbien formalism of general relativity and a theorem about differential constraint

In principle, for removing all second time derivative terms in the ten Einstein equations, we must use tetrad $e^{\hat\alpha}_\mu$ rather than metric tensor $g_{\mu\nu}$ as basic variables. Because if we use metric tensor $g_{\mu\nu}$ as basic variables, then for choosing special form of metric tensor, what we can use are only four gauge conditions provided by coordinate transformation $x^\mu = x^\mu(\widetilde{x}^\nu)$, since metric tensor cannot be changed by local Lorentz transformation $e^{\hat\alpha}_\mu(x) = \Lambda^{\hat\alpha}_{\hat\beta}(x)\widetilde{e}^{\hat\beta}_\mu(x)$; however, there are the six dynamical variables $g_{ij}$, i.e., the six second time derivative terms $g_{ij,0,0}$, in the ten Einstein equations. Generally speaking, four conditions cannot eliminate six terms. On the other hand, if we use tetrad $e^{\hat\alpha}_\mu$ as basic variables, then there are only six dynamical equations including second time derivative terms in the ten Einstein equations yet, but for choosing special form of tetrad, ten gauge conditions can now be used, the four are provided by coordinate transformation $x^\mu = x^\mu(\widetilde{x}^\nu)$ and the six are provided by local Lorentz transformation $e^{\hat\alpha}_\mu(x) = \Lambda^{\hat\alpha}_{\hat\beta}(x)\widetilde{e}^{\hat\beta}_\mu(x)$.

We therefore regard tetrad field $e^{\hat\alpha}_\mu$ as basic variables of gravitational field, in the following discussion, $g_{\mu\nu}$ and $g^{\mu\nu}$ are as the abbreviation for $e^{\hat\alpha}_\mu e_{\hat\alpha\nu}$ and $e^{\hat\alpha\mu}e^\nu_{\hat\alpha}$, respectively.

Substituting $g_{\mu\nu} = e^{\hat\alpha}_\mu e_{\hat\alpha\nu}$ and $g^{\mu\nu} = e^{\hat\alpha\mu}e^\nu_{\hat\alpha}$ to (1-3) and (1-4), we obtain the vierbein forms of $L_{\rm GK}$ and $L_{\rm GU}$ immediately. However, we write $L_{\rm GK}$ given by (1-3) to another form (2-1), which is convenient for the following discussion.

Defining $\bar{e}^i_{\hat a} = e^i_{\hat a} - \dfrac{e^0_{\hat a} e^i_{\hat 0}}{e^0_{\hat 0}}$, which satisfies $\bar{e}^i_{\hat a} e^{\hat a}_j = \delta^i_j$, $\bar{e}^i_{\hat a} e^{\hat b}_i = \delta^{\hat b}_{\hat a}$, $\bar{e}^i_{\hat a}$ is thus the inverse matrix of $e^{\hat a}_i$; and notice that $\widetilde{g}^{ij} e^{\hat a}_i e^{\hat b}_j = \eta^{\hat a\hat b} - \dfrac{e^{\hat a 0} e^{\hat b 0}}{g^{00}}$. Using these formulas, we obtain

$$\begin{aligned} L_{\rm GK} &= \frac{1}{-g^{00}}\left(\widetilde{g}^{pu}\widetilde{g}^{qv} - \widetilde{g}^{pq}\widetilde{g}^{uv}\right) e^{\hat a}_p e^{\hat b}_q e^{\hat c}_u e^{\hat d}_v \left(\bar{e}^i_{\hat a}\bar{e}^j_{\hat b}\Gamma^0_{ij}\bar{e}^l_{\hat c}\bar{e}^m_{\hat d}\Gamma^0_{lm}\right) \\ &= \frac{1}{-g^{00}}\left[\left(\eta^{\hat a\hat c} - \frac{e^{\hat a 0}e^{\hat c 0}}{g^{00}}\right)\left(\eta^{\hat b\hat d} - \frac{e^{\hat b 0}e^{\hat d 0}}{g^{00}}\right) - \left(\eta^{\hat a\hat b} - \frac{e^{\hat a 0}e^{\hat b 0}}{g^{00}}\right)\left(\eta^{\hat c\hat d} - \frac{e^{\hat c 0}e^{\hat d 0}}{g^{00}}\right)\right] Y_{\hat a\hat b} Y_{\hat c\hat d}, \end{aligned} \tag{2-1}$$

where

$$
\begin{aligned}
Y_{\hat{a}\hat{b}} &= \bar{e}^i_{\hat{a}}\bar{e}^j_{\hat{b}}\Gamma^0_{ij} \\
&= -\frac{1}{2}g^{0\lambda}\left[\bar{e}^i_{\hat{a}}\left(e_{\hat{b}i,\lambda}-e_{\hat{b}\lambda,i}\right)+\bar{e}^i_{\hat{b}}\left(e_{\hat{a}i,\lambda}-e_{\hat{a}\lambda,i}\right)\right] \\
&\quad +\frac{1}{2}\bar{e}^i_{\hat{a}}\bar{e}^j_{\hat{b}}e^0_{\hat{\gamma}}\left(e^{\hat{\gamma}}_{i,j}+e^{\hat{\gamma}}_{j,i}\right)-\frac{1}{2}\frac{1}{e^0_{\hat{0}}}g^{0\lambda}\left(e^0_{\hat{a}}\bar{e}^i_{\hat{b}}+e^0_{\hat{b}}\bar{e}^i_{\hat{a}}\right)\left(e^{\hat{0}}_{i,\lambda}-e^{\hat{0}}_{\lambda,i}\right).
\end{aligned}
\tag{2-2}
$$

The vierbein forms of the Einstein equations have been given in Ref. [1].

For removing all quadratic terms of first time derivative of the Lagrangian (2-1), we first prove a theorem.

As well known, when we investigate a system with constraint conditions, an algebraic constraint in which there is not derivative term can be putted directly in the Lagrangian density of the system, but, generally speaking, we cannot do like so for differential constraint including derivative term. However, if a differential constraint appears in quadratic form in the Lagrangian density of a system, then we can prove the following theorem.

**Theorem.** If the Lagrangian density $L(\varphi_a,\varphi_{a,\lambda})$ of a system whose basic variables are $\varphi_a$ $(a=1,2,\cdots,N)$ has the form

$$L(\varphi_a,\varphi_{a,\lambda})=W(\varphi_a,\varphi_{a,\lambda})\left[\chi(\varphi_a,\varphi_{a,\lambda})\right]^2+L'(\varphi_a,\varphi_{a,\lambda}),$$

then if we add the constraint condition $\chi(\varphi_a,\varphi_{a,\lambda})=0$, then the term $W(\varphi_a,\varphi_{a,\lambda})\left[\chi(\varphi_a,\varphi_{a,\lambda})\right]^2$ in $L(\varphi_a,\varphi_{a,\lambda})$ can be removed directly, namely, we can use $L'(\varphi_a,\varphi_{a,\lambda})$ as the Lagrangian density of the field $\varphi_a$.

The proof of the theorem is quite simple. The Euler-Lagrange equation corresponding to $L(\varphi_a,\varphi_{a,\lambda})$ is

$$
\begin{aligned}
&\frac{\partial L(\varphi_b,\varphi_{b,\tau})}{\partial\varphi_a}-\partial_\lambda\frac{\partial L(\varphi_b,\varphi_{b,\tau})}{\partial\varphi_{a,\lambda}} \\
&=\frac{\partial\left\{W(\varphi_b,\varphi_{b,\tau})\left[\chi(\varphi_b,\varphi_{b,\tau})\right]^2\right\}}{\partial\varphi_a}-\partial_\lambda\frac{\partial\left\{W(\varphi_b,\varphi_{b,\tau})\left[\chi(\varphi_b,\varphi_{b,\tau})\right]^2\right\}}{\partial\varphi_{a,\lambda}} \\
&\quad+\frac{\partial L'(\varphi_b,\varphi_{b,\tau})}{\partial\varphi_a}-\partial_\lambda\frac{\partial L'(\varphi_b,\varphi_{b,\tau})}{\partial\varphi_{a,\lambda}} \\
&=\frac{\partial W(\varphi_b,\varphi_{b,\tau})}{\partial\varphi_a}\left[\chi(\varphi_b,\varphi_{b,\tau})\right]^2+2W(\varphi_b,\varphi_{b,\tau})\chi(\varphi_b,\varphi_{b,\tau})\frac{\partial\chi(\varphi_b,\varphi_{b,\tau})}{\partial\varphi_a} \\
&\quad-\partial_\lambda\left\{\frac{\partial W(\varphi_b,\varphi_{b,\tau})}{\partial\varphi_{a,\lambda}}\left[\chi(\varphi_b,\varphi_{b,\tau})\right]^2+2W(\varphi_b,\varphi_{b,\tau})\chi(\varphi_b,\varphi_{b,\tau})\frac{\partial\chi(\varphi_b,\varphi_{b,\tau})}{\partial\varphi_{a,\lambda}}\right\} \\
&\quad+\frac{\partial L'(\varphi_b,\varphi_{b,\tau})}{\partial\varphi_a}-\partial_\lambda\frac{\partial L'(\varphi_b,\varphi_{b,\tau})}{\partial\varphi_{a,\lambda}},
\end{aligned}
$$

if we add the constraint condition $\chi(\varphi_a,\varphi_{a,\lambda})=0$, then the above Euler-Lagrange equation becomes

$$\frac{\partial L(\varphi_b,\varphi_{b,\tau})}{\partial\varphi_a}-\partial_\lambda\frac{\partial L(\varphi_b,\varphi_{b,\tau})}{\partial\varphi_{a,\lambda}}=\frac{\partial L'(\varphi_b,\varphi_{b,\tau})}{\partial\varphi_a}-\partial_\lambda\frac{\partial L'(\varphi_b,\varphi_{b,\tau})}{\partial\varphi_{a,\lambda}}.$$

This means that we can obtain the equation of motion of the field $\varphi_a$ by the Euler-Lagrange equation corresponding to $L'(\varphi_a, \varphi_{a,\lambda})$.

For example, the action of electromagnetic field in flat spacetime is

$$\begin{aligned} S_{\text{EM}} &= \int \mathrm{d}^4 x \left( -\frac{1}{4}(A_{\mu,\nu} - A_{\nu,\mu})(A^{\mu,\nu} - A^{\nu,\mu}) - ej^\mu A_\mu \right) \\ &= \int \mathrm{d}^4 x \left( -\frac{1}{2} A_{\rho,\sigma} A^{\rho,\sigma} + \frac{1}{2}\left(A^\lambda{}_{,\lambda}\right)^2 - ej^\mu A_\mu \right) + \frac{1}{2}\int \mathrm{d}^4 x \frac{\partial\left(A^\rho A^\sigma{}_{,\rho} - A^\sigma A^\rho{}_{,\rho}\right)}{\partial x^\sigma}, \end{aligned} \tag{2-3}$$

where the last term is an integral of a total derivative, which does not impact on the derivation of the equations of motion. If we add the Lorenz gauge condition $A^\lambda{}_{,\lambda} = 0$, then we can remove the quadratic term $\frac{1}{2}\left(A^\lambda{}_{,\lambda}\right)^2$ in (2-3) and, thus, (2-3) becomes

$$\widetilde{S}_{\text{EM}} = \int \mathrm{d}^4 x \left( -\frac{1}{2} A_{\rho,\sigma} A^{\rho,\sigma} - ej^\mu A_\mu \right), \tag{2-4}$$

from which we obtain the equation of motion

$$A^{\mu,\nu}{}_{,\nu} = ej^\mu. \tag{2-5}$$

**2.2 The time gauge condition**

Under local Lorentz transformation, the manner of transformation of $e_\mu^{\hat{\alpha}}$ reads

$$e_\mu^{\hat{\alpha}}(x) = \Lambda_{\hat{\beta}}^{\hat{\alpha}}(x)\tilde{e}_\mu^{\hat{\beta}}(x),\ e_{\hat{\alpha}}^\mu(x) = \overline{\Lambda}_{\hat{\alpha}}^{\hat{\beta}}(x)\tilde{e}_{\hat{\beta}}^\mu(x), \tag{2-6}$$

where $\Lambda_{\hat{\beta}}^{\hat{\gamma}}\overline{\Lambda}_{\hat{\gamma}}^{\hat{\alpha}} = \delta_{\hat{\beta}}^{\hat{\alpha}}$, $\Lambda_{\hat{\gamma}}^{\hat{\alpha}}\overline{\Lambda}_{\hat{\beta}}^{\hat{\gamma}} = \delta_{\hat{\beta}}^{\hat{\alpha}}$ .

We first choose three gauge-fixing terms, so called the Schwinger time gauge condition:

$$e_{\hat{a}}^0 = 0,\ \ a = 1,2,3, \tag{2-7}$$

which can be implemented by choosing appropriate $\Lambda_0^{\hat{a}}(x)$ of $\Lambda_{\hat{\beta}}^{\hat{\alpha}}(x)$ indicated by (2-6). Under the condition (2-7) we have

$$\begin{aligned} &e_i^{\hat{0}} = 0,\ i = 1,2,3;\ e_{\hat{0}}^0 = \left(e_0^{\hat{0}}\right)^{-1},\ e_{\hat{0}}^i = -e_{\hat{0}}^0 e_0^{\hat{a}} e_{\hat{a}}^i,\ e_{\hat{a}}^i e_i^{\hat{b}} = \delta_{\hat{a}}^{\hat{b}},\ e_{\hat{a}}^j e_i^{\hat{a}} = \delta_i^j; \\ &g^{00} = -\left(e_{\hat{0}}^0\right)^2,\ g^{0i} = -e_{\hat{0}}^0 e_{\hat{0}}^i,\ g^{ij} = -e_{\hat{0}}^i e_{\hat{0}}^j + e_{\hat{a}}^i e^{\hat{a}j},\ \tilde{g}^{ij} = e_{\hat{a}}^i e^{\hat{a}j}; \\ &g_{00} = -\left(e_0^{\hat{0}}\right)^2 + e_0^{\hat{a}} e_{\hat{a}0},\ g_{0i} = e_0^{\hat{a}} e_{\hat{a}i},\ g_{ij} = e_i^{\hat{a}} e_{\hat{a}j},\ \sqrt{-g} = \left\|{}^4 e\right\| = e_0^{\hat{0}}\left\|{}^3 e\right\|. \end{aligned} \tag{2-8}$$

where $\left\|{}^4 e\right\| = \det\left[e_\mu^{\hat{\alpha}}\right]$ and $\left\|{}^3 e\right\| = \det\left[e_i^{\hat{a}}\right]$ are the determinants of the 4×4 matrix $\left[e_\mu^{\hat{\alpha}}\right]$ and the 3 ×3 matrix $\left[e_i^{\hat{a}}\right]$, respectively.

Under the time gauge condition (2-7), (2-1) and (2-2) become

$$
\begin{aligned}
L_{\mathrm{GK}} &= \left(e_0^{\hat{0}}\right)^2 \left(\eta^{\hat{a}\hat{c}}\eta^{\hat{b}\hat{d}} - \eta^{\hat{a}\hat{b}}\eta^{\hat{c}\hat{d}}\right) Y_{\hat{a}\hat{b}} Y_{\hat{c}\hat{d}} \\
&= 2\left(e_0^{\hat{0}}\right)^2 \left(-Y_{\hat{1}\hat{1}}Y_{\hat{2}\hat{2}} - Y_{\hat{1}\hat{1}}Y_{\hat{3}\hat{3}} - Y_{\hat{2}\hat{2}}Y_{\hat{3}\hat{3}} + \left(Y_{\hat{1}\hat{2}}\right)^2 + \left(Y_{\hat{2}\hat{3}}\right)^2 + \left(Y_{\hat{3}\hat{1}}\right)^2\right) \\
&= \left(e_0^{\hat{0}}\right)^2 \left(-\frac{2}{3}\left(Y_{\hat{a}}^{\hat{a}}\right)^2 + \frac{1}{6}\left(2Y_{\hat{1}\hat{1}} - Y_{\hat{2}\hat{2}} - Y_{\hat{3}\hat{3}}\right)^2 + \frac{1}{2}\left(Y_{\hat{2}\hat{2}} - Y_{\hat{3}\hat{3}}\right)^2 + 2\left(Y_{\hat{1}\hat{2}}\right)^2 + 2\left(Y_{\hat{2}\hat{3}}\right)^2 + 2\left(Y_{\hat{3}\hat{1}}\right)^2\right),
\end{aligned}
\tag{2-9}
$$

$$
Y_{\hat{a}\hat{b}} = e_{\hat{a}}^i e_{\hat{b}}^j \Gamma_{ij}^0 = \frac{1}{2} e_0^{\hat{0}} e_{\hat{0}}^{\lambda} \left[ e_{\hat{a}}^i \left(e_{\hat{b}i,\lambda} - e_{\hat{b}\lambda,i}\right) + e_{\hat{b}}^i \left(e_{\hat{a}i,\lambda} - e_{\hat{a}\lambda,i}\right)\right], \tag{2-10}
$$

respectively. For $Y_{\hat{a}\hat{b}}$ we have

$$
\begin{aligned}
& Y_{\hat{a}\hat{b}} = Y_{\hat{b}\hat{a}}, \; Y_{\hat{b}}^{\hat{a}} = \eta^{\hat{a}\hat{c}} Y_{\hat{c}\hat{b}}, \; Y^{\hat{a}\hat{b}} = \eta^{\hat{a}\hat{c}}\eta^{\hat{b}\hat{d}} Y_{\hat{c}\hat{d}}, \\
& Y_{\hat{a}}^{\hat{a}} = \eta^{\hat{a}\hat{b}} Y_{\hat{a}\hat{b}} = Y_{\hat{1}\hat{1}} + Y_{\hat{2}\hat{2}} + Y_{\hat{3}\hat{3}} = e_0^{\hat{0}} e_{\hat{0}}^{\lambda} e_{\hat{a}}^i \left(e_{i,\lambda}^{\hat{a}} - e_{\lambda,i}^{\hat{a}}\right) \\
& = \eta^{\hat{a}\hat{b}} e_{\hat{a}}^i e_{\hat{b}}^j \Gamma_{ij}^0 = \tilde{g}^{ij}\Gamma_{ij}^0 = \frac{\sqrt{-g^{00}}}{\sqrt{-g}} \left(\sqrt{\|g_{lm}\|}\,\frac{g^{0\lambda}}{g^{00}}\right)_{,\lambda},
\end{aligned}
\tag{2-11}
$$

notice that $Y_{\hat{a}}^{\hat{a}}$ can be expressed by metric tensor.

According to (1-3), (2-9) and (2-11), we see that under the time gauge condition (2-7), we have

$$
\begin{aligned}
& \tilde{g}^{il}\tilde{g}^{jm}\Gamma_{ij}^0\Gamma_{lm}^0 - \frac{1}{3}\left(\tilde{g}^{il}\Gamma_{ij}^0\right)^2 \\
& = \frac{1}{6}\left(2Y_{\hat{1}\hat{1}} - Y_{\hat{2}\hat{2}} - Y_{\hat{3}\hat{3}}\right)^2 + \frac{1}{2}\left(Y_{\hat{2}\hat{2}} - Y_{\hat{3}\hat{3}}\right)^2 + 2\left(Y_{\hat{1}\hat{2}}\right)^2 + 2\left(Y_{\hat{2}\hat{3}}\right)^2 + 2\left(Y_{\hat{3}\hat{1}}\right)^2 \geq 0.
\end{aligned}
\tag{2-12}
$$

It is obvious that there is a negative kinetic energy term in (2-9), of which the concrete form is

$$
\begin{aligned}
& L_{\mathrm{GNK}} = -\frac{2}{3}\frac{1}{-g^{00}}\left(Y_{\hat{a}}^{\hat{a}}\right)^2 = \frac{2}{3}\frac{1}{g^{00}}\left(\tilde{g}^{ij}\Gamma_{ij}^0\right)^2 = \frac{2}{3g}\left[\left(\sqrt{\|g_{lm}\|}\,\frac{g^{0\lambda}}{g^{00}}\right)_{,\lambda}\right]^2, \\
& \sqrt{-g}\,L_{\mathrm{GNK}} = -\frac{2}{3\sqrt{-g}}\left[\left(\sqrt{\|g_{lm}\|}\,\frac{g^{0\lambda}}{g^{00}}\right)_{,\lambda}\right]^2.
\end{aligned}
\tag{2-13}
$$

The above forms of the negative kinetic energy term have been discussed in detail in Ref. [1].

Using the formulas in (2-7) and (2-8), it can be proved that $L_{\mathrm{GU}}$ given by (1-4) becomes

$$
L_{\mathrm{GU}} = -2e_0^{\hat{0}} e_{0,k}^{\hat{0}} U^k - L_{\mathrm{GV}} + \frac{1}{e_0^{\hat{0}}\left\|{}^3e\right\|}\frac{\partial}{\partial x^i}\left[e_0^{\hat{0}}\left\|{}^3e\right\|\left(e_{\hat{a}}^i\frac{\partial e^{\hat{a}j}}{\partial x^j} - e_{\hat{a}}^j\frac{\partial e^{\hat{a}i}}{\partial x^j}\right)\right], \tag{2-14}
$$

$$
\begin{aligned}
& U^k = e_{\hat{a}}^i e_{\hat{b}}^j e^{\hat{b}k}\left(e_{i,j}^{\hat{a}} - e_{j,i}^{\hat{a}}\right), \\
& L_{\mathrm{GV}} = e^{\hat{c}l} e_{\hat{c}}^m \left(-\frac{1}{4}\eta_{\hat{a}\hat{b}} e^{\hat{d}i} e_{\hat{d}}^j - \frac{1}{2} e_{\hat{a}}^j e_{\hat{b}}^i + e_{\hat{a}}^i e_{\hat{b}}^j\right)\left(e_{i,l}^{\hat{a}} - e_{l,i}^{\hat{a}}\right)\left(e_{j,m}^{\hat{b}} - e_{m,j}^{\hat{b}}\right).
\end{aligned}
\tag{2-15}
$$

Hence, up to a total derivative, we can employ

$$
S_{(1)} = \int e_0^{\hat{0}}\left\|{}^3e\right\| \mathrm{d}^4x L_{\mathrm{Total}(1)}(x), \quad L_{\mathrm{Total}(1)} = \frac{c^3}{16\pi G}\left(L_{\mathrm{GK}} + 2e_0^{\hat{0}} e_{0,k}^{\hat{0}} U^k + L_{\mathrm{GV}}\right) + L_{\mathrm{M}} \tag{2-16}
$$

as the action of the system whose thirteen basic variables are $\left\{e_0^{\hat{\gamma}}, e_i^{\hat{a}}\right\}$, where $L_{\rm GK}$, $U^i$ and $L_{\rm GV}$ are given by (2-9) and (2-15), respectively; $L_{\rm M}(x)$ is the Lagrangian density of matter.

The Euler-Lagrange equations corresponding to the action (2-16) are:

$$\frac{\partial\left[e_0^{\hat{0}}\left\|{}^3e\right\|L_{\rm Total(1)}\right]}{\partial e_0^{\hat{\gamma}}}-\partial_\nu\frac{\partial\left[e_0^{\hat{0}}\left\|{}^3e\right\|L_{\rm Total(1)}\right]}{\partial e_{0,\nu}^{\hat{\gamma}}}=-\frac{c^3\left\|{}^3e\right\|}{8\pi G}e_0^{\hat{0}}\Theta_{\hat{\gamma}}^0=0\,, \tag{2-17}$$

$$\frac{\partial\left[e_0^{\hat{0}}\left\|{}^3e\right\|L_{\rm Total(1)}\right]}{\partial e_i^{\hat{a}}}-\partial_\nu\frac{\partial\left[e_0^{\hat{0}}\left\|{}^3e\right\|L_{\rm Total(1)}\right]}{\partial e_{i,\nu}^{\hat{a}}}=-\frac{c^3\left\|{}^3e\right\|}{8\pi G}e_0^{\hat{0}}\Theta_{\hat{a}}^i=0\,; \tag{2-18}$$

where

$$e_0^{\hat{0}}\Theta_{\hat{0}}^0=\frac{1}{\left\|{}^3e\right\|}\frac{\partial}{\partial x^k}\left(\left\|{}^3e\right\|U^k\right)+\frac{1}{2}\left(L_{\rm GK}-L_{\rm GV}\right)-\frac{8\pi G}{c^4}e_0^{\hat{0}}T_{\hat{0}}^0\,, \tag{2-19}$$

$$e_0^{\hat{0}}\Theta_{\hat{a}}^0=\eta_{\hat{a}\hat{b}}\frac{1}{\left\|{}^3e\right\|}\frac{\partial}{\partial x^k}\left(\left\|{}^3e\right\|S^{\hat{0}k\hat{b}}\right)+e_{\hat{a}}^iS^{\hat{0}j\hat{b}}\left(e_{\hat{b}i,j}-e_{\hat{b}j,i}\right)-\frac{8\pi G}{c^4}e_0^{\hat{0}}T_{\hat{a}}^0\,, \tag{2-20}$$

$$\begin{aligned}e_0^{\hat{0}}\Theta_{\hat{a}}^i=&-\eta_{\hat{a}\hat{b}}\frac{1}{\left\|{}^3e\right\|}\frac{\partial}{\partial x^0}\left(\left\|{}^3e\right\|S^{\hat{0}i\hat{b}}\right)+\eta_{\hat{a}\hat{b}}\frac{1}{\left\|{}^3e\right\|}\frac{\partial}{\partial x^j}\left[\left\|{}^3e\right\|\left(e_0^{\hat{c}}e_{\hat{c}}^jS^{\hat{0}i\hat{b}}-e_0^{\hat{c}}e_{\hat{c}}^iS^{\hat{0}j\hat{b}}\right)\right]\\&-\frac{1}{2}e_0^{\hat{0}}e_{\hat{a}}^iL_{\rm GK}-e_0^{\hat{0}}e_{\hat{0}}^ke_{\hat{a}}^kS^{\hat{0}j\hat{b}}\left(e_{\hat{b}j,k}-e_{\hat{b}k,j}\right)-e_0^{\hat{0}}e_{\hat{a}}^jS^{\hat{0}i\hat{b}}e_0^\lambda\left(e_{\hat{b}j,\lambda}-e_{\hat{b}\lambda,j}\right)\\&+\frac{1}{\left\|{}^3e\right\|}\frac{\partial}{\partial x^j}\left[\left\|{}^3e\right\|\left(e_{0,k}^{\hat{0}}e_{\hat{b}}^k\left(e_{\hat{a}}^ie^{\hat{b}j}-e_{\hat{a}}^je^{\hat{b}i}\right)+\eta_{\hat{a}\hat{b}}e_0^{\hat{0}}\tilde{s}^{ij\hat{b}}\right)\right]\\&+e_{0,j}^{\hat{0}}\left(e_{\hat{a}}^jU^i-e_{\hat{a}}^iU^j\right)+e_{0,j}^{\hat{0}}e_{\hat{b}}^je_{\hat{a}}^l\left(e^{\hat{c}i}e^{\hat{b}m}-e^{\hat{b}i}e^{\hat{c}m}\right)\left(e_{\hat{c}l,m}-e_{\hat{c}m,l}\right)\\&+e_0^{\hat{0}}e_{\hat{a}}^j\tilde{s}^{ik\hat{b}}\left(e_{\hat{b}j,k}-e_{\hat{b}k,j}\right)-\frac{1}{2}e_0^{\hat{0}}e_{\hat{a}}^iL_{\rm GV}-\frac{8\pi G}{c^4}e_0^{\hat{0}}T_{\hat{a}}^i\,;\end{aligned} \tag{2-21}$$

In (2-20) and (2-21),

$$S^{\hat{0}i\hat{a}}=e_0^{\hat{0}}\left(e^{\hat{a}i}Y_{\hat{b}}^{\hat{b}}-e^{\hat{b}i}Y_{\hat{b}}^{\hat{a}}\right)=e_0^{\hat{0}}\left(e^{\hat{a}i}Y_{\hat{b}}^{\hat{b}}-e_{\hat{b}}^iY^{\hat{a}\hat{b}}\right), \tag{2-22}$$

$$\begin{aligned}\tilde{s}^{ij\hat{a}}=&-\frac{1}{2}e_{\hat{b}}^ie^{\hat{b}l}e_{\hat{c}}^je^{\hat{c}m}\left(e_{l,m}^{\hat{a}}-e_{m,l}^{\hat{a}}\right)+\frac{1}{2}e^{\hat{a}l}e^{\hat{b}m}\left(e_{\hat{b}}^ie_{\hat{c}}^j-e_{\hat{b}}^je_{\hat{c}}^i\right)\left(e_{l,m}^{\hat{c}}-e_{m,l}^{\hat{c}}\right)\\&-\left(e^{\hat{a}i}e^{\hat{b}j}-e^{\hat{b}i}e^{\hat{a}j}\right)e_{\hat{b}}^le_{\hat{c}}^m\left(e_{l,m}^{\hat{c}}-e_{m,l}^{\hat{c}}\right);\end{aligned} \tag{2-23}$$

$$\begin{aligned}T_{\hat{\gamma}}^0&=\frac{c}{e_0^{\hat{0}}\left\|{}^3e\right\|}\left[\frac{\partial\left(e_0^{\hat{0}}\left\|{}^3e\right\|L_{\rm M}\right)}{\partial e_0^{\hat{\gamma}}}-\partial_\nu\frac{\partial\left(e_0^{\hat{0}}\left\|{}^3e\right\|L_{\rm M}\right)}{\partial e_{0,\nu}^{\hat{\gamma}}}\right],\\T_{\hat{a}}^i&=\frac{c}{e_0^{\hat{0}}\left\|{}^3e\right\|}\left[\frac{\partial\left(e_0^{\hat{0}}\left\|{}^3e\right\|L_{\rm M}\right)}{\partial e_i^{\hat{a}}}-\partial_\nu\frac{\partial\left(e_0^{\hat{0}}\left\|{}^3e\right\|L_{\rm M}\right)}{\partial e_{i,\nu}^{\hat{a}}}\right].\end{aligned} \tag{2-24}$$

All the action $S_{(1)}$, the Lagrangian $L_{\rm Total(1)}$ and the vierbein forms of the Einstein equations under the time gauge condition given by (2-16) and (2-17) ~ (2-24) respectively have been presented and analyzed in Ref. [1] (some formulas rewritten here are in new forms); for example, in Ref. [1] it has been pointed out that we can only obtain ten independent equations from (2-17) ~ (2-24) since in which there are three identities

$$\left(\Theta^{i}_{\hat{a}}-e^{\hat{0}}_{0}e^{i}_{\hat{0}}\Theta^{0}_{\hat{a}}\right)e^{\hat{a}j}=\left(\Theta^{j}_{\hat{a}}-e^{\hat{0}}_{0}e^{j}_{\hat{0}}\Theta^{0}_{\hat{a}}\right)e^{\hat{a}i}\qquad (i,j=1,2,3;\ i\neq j)\ .$$

From (2-19) and (2-20) we see that there is not any second time derivative term in the four equations (2-17), and from (2-21) we see that in the nine equations (2-18), all second time derivative terms only appear in the term $-\eta_{\hat{a}\hat{b}}\frac{1}{\left\|{}^{3}e\right\|}\frac{\partial}{\partial x^{0}}\left(\left\|{}^{3}e\right\|S^{\hat{0}i\hat{b}}\right)$. Hence, if we can eliminate all first time derivative terms in $S^{\hat{0}i\hat{a}}$ given by (2-22) via appropriate gauge conditions, then there is not any second time derivative term in the all equations (2-17) and (2-18).

**2.3 The simplest constraint conditions**

If both two tetrads $\left\{\tilde{e}^{\hat{\alpha}}_{\mu}\right\}$ and $\left\{e^{\hat{\alpha}}_{\mu}\right\}$ satisfy the time gauge condition, namely, (2-7) holds for $\left\{e^{\hat{\alpha}}_{\mu}\right\}$ and

$$\tilde{e}^{0}_{\hat{a}}=0,\ \ a=1,2,3 \tag{2-25}$$

for $\left\{\tilde{e}^{\hat{\alpha}}_{\mu}\right\}$, then a local Lorentz transformation $\Lambda^{\hat{\alpha}}_{\hat{\beta}}$ between such $\left\{e^{\hat{\gamma}}_{0},e^{\hat{a}}_{i}\right\}$ and $\left\{\tilde{e}^{\hat{\gamma}}_{0},\tilde{e}^{\hat{a}}_{i}\right\}$ has the characteristics:

$$\Lambda^{\hat{0}}_{\hat{0}}(x)=1\ ,\ \Lambda^{\hat{a}}_{\hat{0}}(x)=0\ ,\ \Lambda^{\hat{0}}_{\hat{a}}(x)=0\ ,\ \eta^{\hat{c}\hat{d}}\Lambda^{\hat{a}}_{\hat{c}}\Lambda^{\hat{b}}_{\hat{d}}=\eta^{\hat{a}\hat{b}}\,. \tag{2-26}$$

Under the above special local Lorentz transformation, the relation between $\left\{e^{\hat{\gamma}}_{0},e^{\hat{a}}_{i}\right\}$ and $\left\{\tilde{e}^{\hat{\gamma}}_{0},\tilde{e}^{\hat{a}}_{i}\right\}$ reads

$$e^{\mu}_{\hat{0}}=\tilde{e}^{\mu}_{\hat{0}}=\frac{-g^{0\mu}}{\sqrt{-g^{00}}}\,,\ e^{\hat{0}}_{\mu}=\tilde{e}^{\hat{0}}_{\mu}=\frac{\delta^{0}_{\mu}}{\sqrt{-g^{00}}}\,;\ e^{\hat{a}}_{\mu}=\Lambda^{\hat{a}}_{\hat{b}}(x)\tilde{e}^{\hat{b}}_{\mu}\ ,\ e^{\mu}_{\hat{a}}=\overline{\Lambda}^{\hat{b}}_{\hat{a}}(x)\tilde{e}^{\mu}_{\hat{b}}\,. \tag{2-27}$$

All formulas (2-8) ~ (2-24) hold for the tetrad $\left\{\tilde{e}^{\hat{\gamma}}_{0},\tilde{e}^{\hat{a}}_{i}\right\}$, e.g., corresponding to (2-22) and (2-9) we have

$$S^{\hat{0}i\hat{a}}=\tilde{e}^{\hat{0}}_{0}\tilde{e}^{\hat{a}i}\tilde{Y}^{\hat{b}}_{\hat{b}}-\tilde{e}^{\hat{0}}_{0}\tilde{e}^{i}_{\hat{b}}\tilde{Y}^{\hat{a}\hat{b}}\,, \tag{2-28}$$

$$L_{\rm GK}=\left(\tilde{e}^{\hat{0}}_{0}\right)^{2}\left(-\frac{2}{3}\left(\tilde{Y}^{\hat{a}}_{\hat{a}}\right)^{2}+\frac{1}{6}\left(2\tilde{Y}_{\hat{1}\hat{1}}-\tilde{Y}_{\hat{2}\hat{2}}-\tilde{Y}_{\hat{3}\hat{3}}\right)^{2}+\frac{1}{2}\left(\tilde{Y}_{\hat{2}\hat{2}}-\tilde{Y}_{\hat{3}\hat{3}}\right)^{2}+2\left(\tilde{Y}_{\hat{1}\hat{2}}\right)^{2}+2\left(\tilde{Y}_{\hat{2}\hat{3}}\right)^{2}+2\left(\tilde{Y}_{\hat{3}\hat{1}}\right)^{2}\right), \tag{2-29}$$

respectively; And by the symbol "~" we denote those quantities corresponding to $\left\{\tilde{e}^{\hat{\gamma}}_{0},\tilde{e}^{\hat{a}}_{i}\right\}$, for example, the expression (2-10) of $Y_{\hat{a}\hat{b}}$ corresponds to the tetrad $\left\{e^{\hat{\gamma}}_{0},e^{\hat{a}}_{i}\right\}$, but the expression

$$\begin{aligned}\tilde{Y}_{\hat{a}\hat{b}}&=\tilde{e}^{i}_{\hat{a}}\tilde{e}^{j}_{\hat{b}}\Gamma^{0}_{ij}=\frac{1}{2}\tilde{e}^{0}_{\hat{0}}\tilde{e}^{\lambda}_{\hat{0}}\left[\tilde{e}^{i}_{\hat{a}}\left(\tilde{e}_{\hat{b}i,\lambda}-\tilde{e}_{\hat{b}\lambda,i}\right)+\tilde{e}^{i}_{\hat{b}}\left(\tilde{e}_{\hat{a}i,\lambda}-\tilde{e}_{\hat{a}\lambda,i}\right)\right]\\&=\frac{1}{2}\left(\tilde{e}^{0}_{\hat{0}}\right)^{2}\left\{\left[\tilde{e}^{i}_{\hat{a}}\left(\tilde{e}_{\hat{b}i,0}-\tilde{e}_{\hat{b}0,i}\right)+\tilde{e}^{i}_{\hat{b}}\left(\tilde{e}_{\hat{a}i,0}-\tilde{e}_{\hat{a}0,i}\right)\right]-\tilde{e}^{\hat{c}}_{0}\tilde{e}^{j}_{\hat{c}}\left[\tilde{e}^{i}_{\hat{a}}\left(\tilde{e}_{\hat{b}i,j}-\tilde{e}_{\hat{b}j,i}\right)+\tilde{e}^{i}_{\hat{b}}\left(\tilde{e}_{\hat{a}i,j}-\tilde{e}_{\hat{a}j,i}\right)\right]\right\}\end{aligned} \tag{2-30}$$

of $\widetilde{Y}_{\hat{a}\hat{b}}$ corresponds to the tetrad $\left\{\widetilde{e}_0^{\hat{\gamma}}, \widetilde{e}_i^{\hat{a}}\right\}$.

Because local Lorentz transformation (2-6) cannot change metric tensor $g_{ij}$, according (2-27), the manner of transformation of $Y_{\hat{a}\hat{b}}$ given by (2-10) under the transformation matrix (2-26) is

$$\widetilde{Y}_{\hat{a}\hat{b}} = \widetilde{e}_{\hat{a}}^i \widetilde{e}_{\hat{b}}^j \Gamma_{ij}^0 = \Lambda_{\hat{a}}^{\hat{c}} e_{\hat{c}}^i \Lambda_{\hat{b}}^{\hat{d}} e_{\hat{d}}^j \Gamma_{ij}^0 = \Lambda_{\hat{a}}^{\hat{c}} \Lambda_{\hat{b}}^{\hat{d}} Y_{\hat{c}\hat{d}} \,. \tag{2-31}$$

According to the corresponding mathematical theorem, if the non diagonal elements of the 3 ×3 symmetric matrix $Y_{\hat{a}\hat{b}}$ are not zero, then we can make an orthogonal transformation by an orthogonal matrix $\Lambda_{\hat{a}}^{\hat{b}}$ such that the new 3×3 symmetric matrix $\widetilde{Y}_{\hat{a}\hat{b}}$ is diagonal; Namely, for any tetrad $\left\{e_0^{\hat{\gamma}}, e_i^{\hat{a}}\right\}$ satisfying the time gauge condition (2-7), if $Y_{\hat{a}\hat{b}} \neq 0$ $(a, b = 1, 2, 3;\ a \neq b)$, then we can choose $\Lambda_{\hat{a}}^{\hat{b}}$ such that

$$\widetilde{Y}_{\hat{1}\hat{2}} = 0\,,\ \ \widetilde{Y}_{\hat{2}\hat{3}} = 0\,,\ \ \widetilde{Y}_{\hat{3}\hat{1}} = 0\,, \tag{2-32}$$

for a new group of tetrad $\left\{\widetilde{e}_0^{\hat{\gamma}}, \widetilde{e}_i^{\hat{a}}\right\}$ satisfying the time gauge condition (2-25) yet.

Eq. (2-32) is in fact just so called the simplest constraint conditions in Ref. [1], it originates from *the simplest case* in Ref. [6].

For the new group of tetrad $\left\{\widetilde{e}_0^{\hat{\gamma}}, \widetilde{e}_i^{\hat{a}}\right\}$ that satisfy (2-32), we use (2-32) to eliminate the terms $\widetilde{Y}^{\hat{a}\hat{b}}$ $(b \neq a)$ in $S^{\hat{0}i\hat{a}}$ given by (2-28) and, thus, $S^{\hat{0}i\hat{a}}$ becomes

$$S^{\hat{0}i\hat{a}} = \widetilde{e}_0^{\hat{0}} \widetilde{e}^{\hat{a}i} \widetilde{Y}_{\hat{b}}^{\hat{b}} - \widetilde{e}_0^{\hat{0}} \widetilde{e}^{\hat{a}i} \widetilde{Y}^{\hat{a}\hat{a}} \qquad (a = 1, 2, 3)\,. \tag{2-33}$$

On the other hand, according to the Theorem proved in Sect. 2.1 and (2-32), the term $\left(\widetilde{e}_0^{\hat{0}}\right)^2 \left(2\left(\widetilde{Y}_{\hat{1}\hat{2}}\right)^2 + 2\left(\widetilde{Y}_{\hat{2}\hat{3}}\right)^2 + 2\left(\widetilde{Y}_{\hat{3}\hat{1}}\right)^2\right)$ in (2-29) can be removed directly and, thus, $L_{\mathrm{GK}}$ becomes

$$L_{\mathrm{GK(1)}} = -\frac{2}{3}\left(\widetilde{e}_0^{\hat{0}}\right)^2 \left(\widetilde{Y}_{\hat{a}}^{\hat{a}}\right)^2 + \frac{1}{2}\left(\widetilde{e}_0^{\hat{0}}\right)^2 \left(\frac{1}{3}\left(2\widetilde{Y}_{\hat{1}\hat{1}} - \widetilde{Y}_{\hat{2}\hat{2}} - \widetilde{Y}_{\hat{3}\hat{3}}\right)^2 + \left(\widetilde{Y}_{\hat{2}\hat{2}} - \widetilde{Y}_{\hat{3}\hat{3}}\right)^2\right) . \tag{2-34}$$

So far, all six gauge conditions provided by local Lorentz transformation $e_\mu^{\hat{\alpha}} = \Lambda_{\hat{\beta}}^{\hat{\alpha}}(x) \widetilde{e}_\mu^{\hat{\beta}}$ have run out. If we want to continue to eliminate all first time derivative terms in $S^{\hat{0}i\hat{a}}$ given by (2-33), then we have to employ some coordinate conditions to eliminate all first time derivative terms in the three terms $\widetilde{Y}_{\hat{a}\hat{a}}$ $(a = 1, 2, 3)$ in $S^{\hat{0}i\hat{a}}$. For this purpose, we first find out the expressions in which $\widetilde{Y}_{\hat{a}\hat{a}}$ $(a = 1, 2, 3)$ are expressed by metric tensor.

Since for determining the tetrad $\{\tilde{e}_0^{\hat{\gamma}}, \tilde{e}_i^{\hat{a}}\}$, all six gauge conditions provided by local Lorentz transformation $e_\mu^{\hat{\alpha}} = \Lambda_{\hat{\beta}}^{\hat{\alpha}}(x)\tilde{e}_\mu^{\hat{\beta}}$ have run out, all elements of the transformation matrix $\Lambda_{\hat{\beta}}^{\hat{\alpha}}(x)$ have been determined fully corresponding to a given tetrad, and the tetrad $\{\tilde{e}_0^{\hat{\gamma}}, \tilde{e}_i^{\hat{a}}\}$ has been determined fully corresponding to metric tensor $g_{\mu\nu}$. Concretely, according to $\tilde{g}^{ij} = \tilde{e}_{\hat{a}}^i \tilde{e}^{\hat{a}j}$ given in (2-8) and (2-32) we can determine nine quantities

$$\tilde{e}_{\hat{a}}^i = \tilde{e}_{\hat{a}}^i(g_{\mu\nu}\ ;\ g_{0i,j}, g_{ij,\tau}) \tag{2-35}$$

by solving the nine algebraic equations

$$\tilde{e}_{\hat{a}}^i \tilde{e}^{\hat{a}j} = \tilde{g}^{ij}\ ;\ \tilde{e}_{\hat{1}}^i \tilde{e}_{\hat{2}}^j \Gamma_{ij}^0 = 0,\ \tilde{e}_{\hat{3}}^i \tilde{e}_{\hat{1}}^j \Gamma_{ij}^0 = 0,\ \tilde{e}_{\hat{2}}^i \tilde{e}_{\hat{3}}^j \Gamma_{ij}^0 = 0. \tag{2-36}$$

Notice that according to (2-27) we have had $\tilde{e}_\mu^{\hat{0}}$ and $\tilde{e}_{\hat{0}}^\mu$ expressed by $g_{\mu\nu}$, hence, after we have the expression (2-35), it is easy to obtain the expressions $\tilde{e}_i^{\hat{a}} = \tilde{e}_i^{\hat{a}}(g_{\mu\nu}\ ;\ g_{0i,j}, g_{ij,\tau})$ and $\tilde{e}_0^{\hat{a}} = \tilde{e}_0^{\hat{a}}(g_{\mu\nu}\ ;\ g_{0i,j}, g_{ij,\tau})$, since $\tilde{e}_i^{\hat{a}}$ is the inverse matrix of $\tilde{e}_{\hat{a}}^i$ and according to (2-8) we have $\tilde{e}_0^{\hat{a}} = -\tilde{e}_0^{\hat{0}} \tilde{e}_{\hat{0}}^i \tilde{e}_i^{\hat{a}}$.

In principle, we can obtain the expression $\tilde{Y}_{\hat{a}\hat{a}} = \tilde{Y}_{\hat{a}\hat{a}}(g_{\mu\nu}\ ;\ g_{0i,j}, g_{ij,\tau})\quad (a = 1,2,3)$ by substituting (2-35) to $\tilde{Y}_{\hat{a}\hat{a}} = \tilde{e}_{\hat{a}}^i \tilde{e}_{\hat{a}}^j \Gamma_{ij}^0 \quad (a = 1,2,3)$. However, the quite complex process obtaining (2-35) by solving the nine algebraic equations in (2-36) forces us to use a different method to determine $\tilde{Y}_{\hat{a}\hat{a}} \quad (a = 1,2,3)$.

According to the corresponding mathematical theorem, for determining the orthogonal matrix $\Lambda_{\hat{a}}^{\hat{b}}$, we first solve the equation

$$\left\| Y_{\hat{a}\hat{b}} - \lambda I \right\| = -\lambda^3 + b_1\lambda^2 - b_2\lambda + b_3 = 0 \tag{2-37}$$

of eigenvalues of 3×3 symmetric matrix $Y_{\hat{a}\hat{b}}$. The coefficients of the above cubic equation are

$$\begin{aligned}
b_1 &= Y_{\hat{1}\hat{1}} + Y_{\hat{2}\hat{2}} + Y_{\hat{3}\hat{3}} = Y_{\hat{a}}^{\hat{a}} = \tilde{g}^{ij}\Gamma_{ij}^0\,,\\
b_2 &= Y_{\hat{1}\hat{1}}Y_{\hat{2}\hat{2}} + Y_{\hat{1}\hat{1}}Y_{\hat{3}\hat{3}} + Y_{\hat{2}\hat{2}}Y_{\hat{3}\hat{3}} - \left(Y_{\hat{1}\hat{2}}\right)^2 - \left(Y_{\hat{2}\hat{3}}\right)^2 - \left(Y_{\hat{3}\hat{1}}\right)^2 = \frac{g^{00}}{2}L_{\text{GK}} = -\frac{1}{2}\left(\tilde{g}^{il}\tilde{g}^{jm} - \tilde{g}^{ij}\tilde{g}^{lm}\right)\Gamma_{ij}^0\Gamma_{lm}^0\,,\\
b_3 &= \left\|Y_{\hat{a}\hat{b}}\right\| = \left\|e_{\hat{a}}^i e_{\hat{b}}^j \Gamma_{ij}^0\right\| = \left\|e_{\hat{a}}^i\right\|^2 \left\|\Gamma_{ij}^0\right\| = \left\|\tilde{g}^{lm}\right\| \left\|\Gamma_{ij}^0\right\| = \frac{\left\|\Gamma_{ij}^0\right\|}{\left\|g_{ij}\right\|}\,,
\end{aligned} \tag{2-38}$$

where $\left\|\Gamma_{ij}^0\right\|$ means the determinant of the 3×3 matrix $\left[\Gamma_{ij}^0\right]$. In the expression of $b_2$, we have used (2-9).

Although the formulas for the roots of cubic equation are well known, we are not going to write out the concrete forms of the three roots $\lambda_{(a)}$ $(a=1,2,3)$ of the cubic equation (2-37) here and only point out the following three characteristics about $\lambda_{(a)}$ $(a=1,2,3)$.

① According to the corresponding mathematical theorem, all the three roots $\lambda_{(a)}$ $(a=1,2,3)$ of (2-37) are real;

② From (2-38) we see that all the coefficients of the cubic equation (2-37) have been expressed as functions of metric tensor $g_{\mu\nu}$, $g_{0i,j}$ and $g_{ij,\tau}$, all the three roots $\lambda_{(a)}$ $(a=1,2,3)$ of (2-37) thus as well:

$$\lambda_{(a)} \equiv \lambda_{(a)}(g_{\mu\nu}\,;\; g_{0i,j}\,, g_{ij,\tau}) \quad (a=1,2,3)\,. \tag{2-39}$$

③ According to Viète's formulas, we have

$$b_1 = \lambda_{(1)} + \lambda_{(2)} + \lambda_{(3)}\,,\; b_2 = \lambda_{(1)}\lambda_{(2)} + \lambda_{(1)}\lambda_{(3)} + \lambda_{(2)}\lambda_{(3)}\,,\; b_3 = \lambda_{(1)}\lambda_{(2)}\lambda_{(3)}\,. \tag{2-40}$$

After obtaining the three real roots $\lambda_{(a)}$ $(a=1,2,3)$ of (2-37), we can obtain the corresponding three eigenvectors $U_{(a)}$ $(a=1,2,3)$ determined by the equation $\left[Y_{\hat{a}\hat{b}}\right] U_{(c)} = \lambda_{(c)} U_{(c)}$ $(c=1,2,3)$, where all $U_{(a)}$ $(a=1,2,3)$ are 3 × 1 matrices and satisfy $U_{(a)}^{\mathrm{T}} U_{(b)} = \delta_{(a)(b)}$ $(a,b=1,2,3)$.

Taking advantage of $U_{(a)}$ $(a=1,2,3)$, we can construct the orthogonal matrix $\Lambda_{\hat{a}}^{\hat{b}} = \left[U_{(1)} \quad U_{(2)} \quad U_{(3)}\right]_{ba}$, the matrix form of (2-31) is

$$\left[\widetilde{Y}_{\hat{a}\hat{b}}\right] = \begin{bmatrix} \widetilde{Y}_{\hat{1}\hat{1}} & \widetilde{Y}_{\hat{1}\hat{2}} & \widetilde{Y}_{\hat{1}\hat{3}} \\ \widetilde{Y}_{\hat{2}\hat{1}} & \widetilde{Y}_{\hat{2}\hat{2}} & \widetilde{Y}_{\hat{2}\hat{3}} \\ \widetilde{Y}_{\hat{3}\hat{1}} & \widetilde{Y}_{\hat{3}\hat{2}} & \widetilde{Y}_{\hat{3}\hat{3}} \end{bmatrix} = \Lambda^{\mathrm{T}} \left[Y_{\hat{c}\hat{d}}\right] \Lambda = \begin{bmatrix} U_{(1)}^{\mathrm{T}} \\ U_{(2)}^{\mathrm{T}} \\ U_{(3)}^{\mathrm{T}} \end{bmatrix} \left[Y_{\hat{c}\hat{d}}\right] \left[U_{(1)} \quad U_{(2)} \quad U_{(3)}\right]$$

$$= \begin{bmatrix} U_{(1)}^{\mathrm{T}} \\ U_{(2)}^{\mathrm{T}} \\ U_{(3)}^{\mathrm{T}} \end{bmatrix} \left[\lambda_{(1)} U_{(1)} \quad \lambda_{(2)} U_{(2)} \quad \lambda_{(3)} U_{(3)}\right] = \begin{bmatrix} \lambda_{(1)} & 0 & 0 \\ 0 & \lambda_{(2)} & 0 \\ 0 & 0 & \lambda_{(3)} \end{bmatrix}.$$

The above result not only verifies (2-32) again, but also presents the following conclusion:

$$\widetilde{Y}_{\hat{a}\hat{a}} = \tilde{e}_{\hat{a}}^{i} \tilde{e}_{\hat{a}}^{j} \Gamma_{ij}^{0} = \lambda_{(a)} \equiv \lambda_{(a)}(g_{\mu\nu}\,;\; g_{0i,j}\,, g_{ij,\tau}) \quad (a=1,2,3)\,, \tag{2-41}$$

namely, $\widetilde{Y}_{\hat{a}\hat{a}}$ $(a=1,2,3)$ are just the three roots of the cubic equation (2-37).

**2.4 A coordinate condition eliminating the negative kinetic energy term in $L_{\mathrm{GK}}$**

Generally speaking, coordinate transformation can provide four coordinate conditions, and from (2-41) we see that all $\widetilde{Y}_{\hat{a}\hat{a}}$ $(a=1,2,3)$ are functions of $g_{\mu\nu}$, $g_{0i,j}$ and $g_{ij,\lambda}$ and independent with second derivative $g_{\mu\nu,\rho,\sigma}$, we therefore can choose special combinations of $\widetilde{Y}_{\hat{a}\hat{a}}$ $(a=1,2,3)$ as coordinate conditions.

We first choose

$$\widetilde{Y}_{\hat{a}}^{\hat{a}} = \widetilde{Y}_{\hat{1}\hat{1}} + \widetilde{Y}_{\hat{2}\hat{2}} + \widetilde{Y}_{\hat{3}\hat{3}} = \widetilde{g}^{ij}\Gamma_{ij}^{0} = \frac{\sqrt{-g^{00}}}{\sqrt{-g}}\left(\sqrt{\|g_{lm}\|}\frac{g^{0\lambda}}{g^{00}}\right)_{,\lambda} = 0 \tag{2-42}$$

as one of four coordinate conditions.

After adding the coordinate condition (2-42), (2-33) becomes

$$S^{\hat{0}i\hat{a}} = -\widetilde{e}_{0}^{\hat{0}}\widetilde{e}^{\hat{a}i}\widetilde{Y}^{\hat{a}\hat{a}} \qquad (a=1,2,3)\,; \tag{2-43}$$

The special meaning of the coordinate condition (2-42) is that it can eliminate the negative kinetic energy term expressed by (2-13) in $L_{\mathrm{GK}}$. In fact, according to the Theorem proved in Sect. 2.1 and considering the condition (2-42), the negative kinetic energy term $-\frac{2}{3}\left(e_{0}^{\hat{0}}\right)^{2}\left(\widetilde{Y}_{\hat{a}}^{\hat{a}}\right)^{2}$ in (2-29) or (2-34) can be removed and, thus, $L_{\mathrm{GK}(1)}$ given by (2-34) becomes

$$L_{\mathrm{GK}(2)} = \frac{1}{2}\left(\widetilde{e}_{0}^{\hat{0}}\right)^{2}\left(\frac{1}{3}\left(2\widetilde{Y}_{\hat{1}\hat{1}} - \widetilde{Y}_{\hat{2}\hat{2}} - \widetilde{Y}_{\hat{3}\hat{3}}\right)^{2} + \left(\widetilde{Y}_{\hat{2}\hat{2}} - \widetilde{Y}_{\hat{3}\hat{3}}\right)^{2}\right). \tag{2-44}$$

So far, we have used seven gauge conditions, the six are provided by local Lorentz transformation $e_{\mu}^{\hat{\alpha}} = \Lambda_{\hat{\beta}}^{\hat{\alpha}}(x)\widetilde{e}_{\mu}^{\hat{\beta}}$, the one is a coordinate condition provided by coordinate transformation $x^{\mu} = x^{\mu}(\widetilde{x}^{\nu})$. All these seven gauge conditions have been given in Ref. [1].

**2.5 Two coordinate conditions**

We further choose the following two coordinate conditions

$$2\widetilde{Y}_{\hat{1}\hat{1}} - \widetilde{Y}_{\hat{2}\hat{2}} - \widetilde{Y}_{\hat{3}\hat{3}} = -g^{00}\Phi_{1}, \tag{2-45}$$

$$\widetilde{Y}_{\hat{2}\hat{2}} - \widetilde{Y}_{\hat{3}\hat{3}} = -g^{00}\Phi_{2}, \tag{2-46}$$

where both $\Phi_1$ and $\Phi_2$ are functions of $g_{\mu\nu}$ and first space derivative $g_{\mu\nu,k}$, whose concrete forms will be determined in the following discussion.

Considering (2-41), the three coordinate conditions (2-42), (2-45) and (2-46) can be written to the following forms:

$$\widetilde{Y}_{\hat{1}\hat{1}} = \lambda_{(1)} = -\frac{1}{3}g^{00}\Phi_{1}\,,\ \widetilde{Y}_{\hat{2}\hat{2}} = \lambda_{(2)} = \frac{1}{2}g^{00}\left(\frac{1}{3}\Phi_{1} - \Phi_{2}\right),\ \widetilde{Y}_{\hat{3}\hat{3}} = \lambda_{(3)} = \frac{1}{2}g^{00}\left(\frac{1}{3}\Phi_{1} + \Phi_{2}\right); \tag{2-47}$$

And, further, according to (2-38) and (2-40), the above three coordinate conditions are equivalent to

$$\tilde{g}^{ij}\Gamma^0_{ij}=0\,,\ \frac{\tilde{g}^{il}\tilde{g}^{jm}\Gamma^0_{ij}\Gamma^0_{lm}}{\left(-g^{00}\right)^2}=\frac{1}{2}\left(\frac{1}{3}\Phi_1^2+\Phi_2^2\right),\quad \frac{\left\|\Gamma^0_{ij}\right\|}{\left(-g^{00}\right)^3\left\|g_{ij}\right\|}=\frac{1}{12}\Phi_1\left(\frac{1}{9}\Phi_1^2-\Phi_2^2\right). \qquad (2\text{-}48)$$

Notice that the second of the above three coordinate conditions asks $\tilde{g}^{il}\tilde{g}^{jm}\Gamma^0_{ij}\Gamma^0_{lm}\geq 0$, but this inequality is only a corollary of (2-12).

Under the three coordinate conditions (2-47) and considering $g^{00}=-\left(\tilde{e}^0_{\hat{0}}\right)^2$ given in (2-8), $S^{\hat{0}i\hat{a}}$ given by (2-43) becomes

$$S^{\hat{0}i\hat{1}}=-\frac{1}{3}\tilde{e}^0_{\hat{0}}\tilde{e}^{\hat{1}i}\Phi_1\,,\ S^{\hat{0}i\hat{2}}=\frac{1}{2}\tilde{e}^0_{\hat{0}}\tilde{e}^{\hat{2}i}\left(\frac{1}{3}\Phi_1-\Phi_2\right),\ S^{\hat{0}i\hat{3}}=\frac{1}{2}\tilde{e}^0_{\hat{0}}\tilde{e}^{\hat{3}i}\left(\frac{1}{3}\Phi_1+\Phi_2\right); \qquad (2\text{-}49)$$

$L_{\mathrm{GK(2)}}$ given by (2-44) can be written to the form

$$L_{\mathrm{GK(2)}}=\frac{1}{6}\left(\tilde{e}^{\hat{0}}_0\right)^2\left(2\tilde{Y}_{\hat{1}\hat{1}}-\tilde{Y}_{\hat{2}\hat{2}}-\tilde{Y}_{\hat{3}\hat{3}}+g^{00}\Phi_1\right)^2+\frac{1}{2}\left(\tilde{e}^{\hat{0}}_0\right)^2\left(\tilde{Y}_{\hat{2}\hat{2}}-\tilde{Y}_{\hat{3}\hat{3}}+g^{00}\Phi_2\right)^2$$
$$-\left(\tilde{e}^{\hat{0}}_0\right)^2\left(\frac{1}{3}\left(2\tilde{Y}_{\hat{1}\hat{1}}-\tilde{Y}_{\hat{2}\hat{2}}-\tilde{Y}_{\hat{3}\hat{3}}\right)g^{00}\Phi_1+\left(\tilde{Y}_{\hat{2}\hat{2}}-\tilde{Y}_{\hat{3}\hat{3}}\right)g^{00}\Phi_2\right)-\frac{1}{2}\left(\tilde{e}^{\hat{0}}_0\right)^2\left(\frac{1}{3}\left(g^{00}\Phi_1\right)^2+\left(g^{00}\Phi_2\right)^2\right),$$

according to the Theorem proved in Sect. 2.1 and considering (2-45) and (2-46), both the terms $\frac{1}{6}\left(\tilde{e}^{\hat{0}}_0\right)^2\left(2\tilde{Y}_{\hat{1}\hat{1}}-\tilde{Y}_{\hat{2}\hat{2}}-\tilde{Y}_{\hat{3}\hat{3}}+g^{00}\Phi_1\right)^2$ and $\frac{1}{2}\left(\tilde{e}^{\hat{0}}_0\right)^2\left(\tilde{Y}_{\hat{2}\hat{2}}-\tilde{Y}_{\hat{3}\hat{3}}+g^{00}\Phi_2\right)^2$ in the above expression can be removed directly and, thus, $L_{\mathrm{GK(2)}}$ becomes

$$L_{\mathrm{GK(3)}}=\frac{1}{3}\left(2\tilde{Y}_{\hat{1}\hat{1}}-\tilde{Y}_{\hat{2}\hat{2}}-\tilde{Y}_{\hat{3}\hat{3}}\right)\Phi_1+\left(\tilde{Y}_{\hat{2}\hat{2}}-\tilde{Y}_{\hat{3}\hat{3}}\right)\Phi_2-\frac{1}{2}\left(\tilde{e}^0_{\hat{0}}\right)^2\left(\frac{1}{3}\Phi_1^2+\Phi_2^2\right)$$
$$=\left(\tilde{e}^0_{\hat{0}}\right)^2\frac{1}{3}\Big\{\Big[2\tilde{e}^i_{\hat{1}}\left(\tilde{e}_{\hat{1}i,0}-\tilde{e}_{\hat{1}0,i}\right)-\tilde{e}^i_{\hat{2}}\left(\tilde{e}_{\hat{2}i,0}-\tilde{e}_{\hat{2}0,i}\right)-\tilde{e}^i_{\hat{3}}\left(\tilde{e}_{\hat{3}i,0}-\tilde{e}_{\hat{3}0,i}\right)\Big]$$
$$-\Big[2\tilde{e}^{\hat{c}}_0\tilde{e}^j_{\hat{c}}\tilde{e}^i_{\hat{1}}\left(\tilde{e}_{\hat{1}i,j}-\tilde{e}_{\hat{1}j,i}\right)-\tilde{e}^{\hat{c}}_0\tilde{e}^j_{\hat{c}}\tilde{e}^i_{\hat{2}}\left(\tilde{e}_{\hat{2}i,j}-\tilde{e}_{\hat{2}j,i}\right)-\tilde{e}^{\hat{c}}_0\tilde{e}^j_{\hat{c}}\tilde{e}^i_{\hat{3}}\left(\tilde{e}_{\hat{3}i,j}-\tilde{e}_{\hat{3}j,i}\right)\Big]\Big\}\Phi_1 \qquad (2\text{-}50)$$
$$+\left(\tilde{e}^0_{\hat{0}}\right)^2\Big\{\Big[\tilde{e}^i_{\hat{2}}\left(\tilde{e}_{\hat{2}i,0}-\tilde{e}_{\hat{2}0,i}\right)-\tilde{e}^i_{\hat{3}}\left(\tilde{e}_{\hat{3}i,0}-\tilde{e}_{\hat{3}0,i}\right)\Big]-\Big[\tilde{e}^{\hat{c}}_0\tilde{e}^j_{\hat{c}}\tilde{e}^i_{\hat{2}}\left(\tilde{e}_{\hat{2}i,j}-\tilde{e}_{\hat{2}j,i}\right)-\tilde{e}^{\hat{c}}_0\tilde{e}^j_{\hat{c}}\tilde{e}^i_{\hat{3}}\left(\tilde{e}_{\hat{3}i,j}-\tilde{e}_{\hat{3}j,i}\right)\Big]\Big\}\Phi_2$$
$$-\frac{1}{2}\left(\tilde{e}^0_{\hat{0}}\right)^2\left(\frac{1}{3}\Phi_1^2+\Phi_2^2\right).$$

$L_{\mathrm{GK(3)}}$ can be written to the form

$$L_{\mathrm{GK(3)}}=\frac{1}{3}\left(2\tilde{Y}_{\hat{1}\hat{1}}-\tilde{Y}_{\hat{2}\hat{2}}-\tilde{Y}_{\hat{3}\hat{3}}+g^{00}\Phi_1\right)\Phi_1+\left(\tilde{Y}_{\hat{2}\hat{2}}-\tilde{Y}_{\hat{3}\hat{3}}+g^{00}\Phi_2\right)\Phi_2+\frac{1}{2}\left(\tilde{e}^0_{\hat{0}}\right)^2\left(\frac{1}{3}\Phi_1^2+\Phi_2^2\right),$$

if $L_{\mathrm{GK(3)}}$ appears in Lagrangian density, then although according to (2-45) and (2-46) we have $2\tilde{Y}_{\hat{1}\hat{1}}-\tilde{Y}_{\hat{2}\hat{2}}-\tilde{Y}_{\hat{3}\hat{3}}+g^{00}\Phi_1=0$ and $\tilde{Y}_{\hat{2}\hat{2}}-\tilde{Y}_{\hat{3}\hat{3}}+g^{00}\Phi_2=0$, neither of $\frac{1}{3}\left(2\tilde{Y}_{\hat{1}\hat{1}}-\tilde{Y}_{\hat{2}\hat{2}}-\tilde{Y}_{\hat{3}\hat{3}}+g^{00}\Phi_1\right)\Phi_1$ and $\frac{1}{3}\left(\tilde{Y}_{\hat{2}\hat{2}}-\tilde{Y}_{\hat{3}\hat{3}}+g^{00}\Phi_2\right)\Phi_2$ can be removed from $L_{\mathrm{GK(3)}}$, since both $2\tilde{Y}_{\hat{1}\hat{1}}-\tilde{Y}_{\hat{2}\hat{2}}-\tilde{Y}_{\hat{3}\hat{3}}+g^{00}\Phi_1$ and $\tilde{Y}_{\hat{2}\hat{2}}-\tilde{Y}_{\hat{3}\hat{3}}+g^{00}\Phi_2$ appear in linear but not quadratic forms in $L_{\mathrm{GK(3)}}$. On the other hand, if $L_{\mathrm{GK}(3)}$ appears in equation of motion, for example, the term $\frac{1}{2}L_{\mathrm{GK}(3)}$ in the equations $e_0^{\hat{0}}\Theta_{\hat{0}}^0 = 0$ and $e_0^{\hat{0}}\Theta_{\hat{a}}^i = 0$ given by (2-17) ~ (2-24), then the both terms $\frac{1}{3}\left(2\widetilde{Y}_{\hat{1}\hat{1}} - \widetilde{Y}_{\hat{2}\hat{2}} - \widetilde{Y}_{\hat{3}\hat{3}} + g^{00}\Phi_1\right)\Phi_1$ and $\frac{1}{3}\left(\widetilde{Y}_{\hat{2}\hat{2}} - \widetilde{Y}_{\hat{3}\hat{3}} + g^{00}\Phi_2\right)\Phi_2$ in $L_{\mathrm{GK}(3)}$ can be eliminated immediately and, thus, the term $\frac{1}{2}L_{\mathrm{GK}(3)}$ in (2-19) and (2-21) becomes $\frac{1}{4}\left(\widetilde{e}_{\hat{0}}^{\,0}\right)^2\left(\frac{1}{3}\Phi_1^2 + \Phi_2^2\right)$.

From (2-49) and (2-50) we see that if both $\Phi_1$ and $\Phi_2$ are functions of $g_{\mu\nu}$ and $g_{\mu\nu,k}$, in other words which are independent with time derivative $g_{\mu\nu,0}$, then after which are expressed by tetrad $\left\{\widetilde{e}_0^{\hat{\gamma}}, \widetilde{e}_i^{\hat{a}}\right\}$ via $g_{00} = \widetilde{e}_0^{\hat{\gamma}}\widetilde{e}_{\hat{\gamma}0}$, $g_{0i} = \widetilde{e}_0^{\hat{a}}\widetilde{e}_{\hat{a}i}$ and $g_{ij} = \widetilde{e}_i^{\hat{a}}\widetilde{e}_{\hat{a}j}$, there is not any time derivative $\widetilde{e}_{0,0}^{\hat{\gamma}}$ or $\widetilde{e}_{i,0}^{\hat{a}}$ in $S^{\hat{0}i\hat{a}}$ given by (2-49) and there is not any quadratic term of first time derivative in $L_{\mathrm{GK}(3)}$ given by (2-50). And, further, since the term $\frac{1}{2}L_{\mathrm{GK}}$ in the equation $e_0^{\hat{0}}\Theta_{\hat{0}}^0 = 0$ and $e_0^{\hat{0}}\Theta_{\hat{a}}^i = 0$ is replaced with $\frac{1}{4}\left(\widetilde{e}_{\hat{0}}^{\,0}\right)^2\left(\frac{1}{3}\Phi_1^2 + \Phi_2^2\right)$, we see that there is not any time derivative term in the equations of motion given by (2-17), (2-19) and (2-20), and there is not any second time derivative term in the equations of motion given by (2-18) and (2-21).

We emphasize again that there is not any time derivative term in $S^{\hat{0}i\hat{a}}$ only when $S^{\hat{0}i\hat{a}}$ given by (2-49) is expressed by tetrad, if it is expressed by metric tensor, then even if both $\Phi_1$ and $\Phi_2$ are independent with time derivative terms $g_{\mu\nu,0}$, from (2-35) we see that $\widetilde{e}^{\hat{a}i}$ in the expression (2-49) of $S^{\hat{0}i\hat{a}}$ is dependent with time derivative terms $g_{ij,0}$, $S^{\hat{0}i\hat{a}}$ thus as well.

Furthermore, notice

$$\Gamma_{ij}^0 = -\frac{1}{2}g^{00}\left[g_{ij,0} + \frac{g^{0k}}{g^{00}}g_{ij,k} + \left(\frac{g^{0k}}{g^{00}}\right)_{,i} g_{kj} + \left(\frac{g^{0k}}{g^{00}}\right)_{,j} g_{ki}\right], \tag{2-51}$$

if we choose both $\Phi_1$ and $\Phi_2$ are functions of $\frac{g^{0i}}{g^{00}}$, $g_{lm}$, $\left(\frac{g^{0i}}{g^{00}}\right)_{,j}$ and $g_{lm,n}$:

$$\Phi_1 \equiv \Phi_1\left(\frac{g^{0i}}{g^{00}},\, g_{lm};\, \left(\frac{g^{0i}}{g^{00}}\right)_{,j},\, g_{lm,n}\right),\quad \Phi_2 \equiv \Phi_2\left(\frac{g^{0i}}{g^{00}},\, g_{lm};\, \left(\frac{g^{0i}}{g^{00}}\right)_{,j},\, g_{lm,n}\right), \tag{2-52}$$

then in the three coordinate conditions (2-48), $g^{00}$ only appear in the combined-item $\frac{g^{0i}}{g^{00}}$ but not alone; And, further, the three coordinate conditions in (2-48) are consistent and independent, since we can regard them as three equations determining three functions $\frac{g^{0i}}{g^{00}}$.

Substituting $g^{00} = -\left(\tilde{e}_0^{\hat{0}}\right)^2$ and $g^{0i} = \left(\tilde{e}_{\hat{0}}^{0}\right)^2 \tilde{e}_0^{\hat{a}} \tilde{e}_{\hat{a}}^{i}$ obtained by (2-8) and $g_{ij} = \tilde{e}_i^{\hat{a}} \tilde{e}_{\hat{a}j}$ to (2-52), we have

$$\Phi_1 \equiv \Phi_1\left(\tilde{e}_0^{\hat{a}} \tilde{e}_{\hat{a}}^{i},\ \tilde{e}_l^{\hat{a}} \tilde{e}_{\hat{a}m};\ \left(\tilde{e}_0^{\hat{a}} \tilde{e}_{\hat{a}}^{i}\right)_{,j},\ \left(\tilde{e}_l^{\hat{a}} \tilde{e}_{\hat{a}m}\right)_{,n}\right),\quad \Phi_2 \equiv \Phi_2\left(\tilde{e}_0^{\hat{a}} \tilde{e}_{\hat{a}}^{i},\ \tilde{e}_l^{\hat{a}} \tilde{e}_{\hat{a}m};\ \left(\tilde{e}_0^{\hat{a}} \tilde{e}_{\hat{a}}^{i}\right)_{,j},\ \left(\tilde{e}_l^{\hat{a}} \tilde{e}_{\hat{a}m}\right)_{,n}\right), \tag{2-53}$$

we see that both $\Phi_1$ and $\Phi_2$ are independent with $\tilde{e}_0^{\hat{0}}$.

Not both $\Phi_1$ and $\Phi_2$ vanish, or from (2-49) we have $S^{\hat{0}i\hat{a}} = 0$ and, according to (2-17) and (2-20), we obtain a contradictory equation $e_0^{\hat{0}} \Theta_{\hat{a}}^{0} = -\frac{8\pi G}{c^4} T_{\hat{a}}^{\hat{0}} = 0$.

A simple choose of $\Phi_1$ and $\Phi_2$ is

$$\Phi_1 = -\left(\frac{g^{0i}}{g^{00}}\right)_{,i} = \left(\tilde{e}_0^{\hat{a}} \tilde{e}_{\hat{a}}^{i}\right)_{,i},\quad \Phi_2 = 0, \tag{2-54}$$

a different choose is

$$\Phi_1 = a\frac{g^{0i}}{g^{00}} \tilde{g}^{jk} g_{ij,k} = -a\tilde{e}_0^{\hat{a}} \tilde{e}_{\hat{a}}^{i} \tilde{e}_{\hat{b}}^{j} \tilde{e}^{\hat{b}k} \left(\tilde{e}_i^{\hat{c}} \tilde{e}_{\hat{c}j}\right)_{,k},\quad \Phi_2 = -b\left(\frac{g^{0i}}{g^{00}}\right)_{,i} = b\left(\tilde{e}_0^{\hat{a}} \tilde{e}_{\hat{a}}^{i}\right)_{,i}, \tag{2-55}$$

where both *a* and *b* are constants, etc. When $\Phi_1$ and $\Phi_2$ are given by (2-54) or (2-55), the three coordinate conditions in (2-48) hold for the metric tensor given by (1-7). Of course, one can try to choose other forms of $\Phi_1$ and $\Phi_2$.

**2.6 General relativity as a fully singular Lagrange system**

We first summarize some characteristics of the group of tetrad $\left\{\tilde{e}_\mu^{\hat{\alpha}}\right\}$ given in the above discussion.

At first, we rewrite the nine gauge conditions given by (2-25), (2-32) and (2-47) as follows.

$$\tilde{e}_{\hat{a}}^{0} = 0\quad (a = 1,2,3); \tag{2-56}$$

$$\left(\tilde{e}_0^{\hat{0}}\right)^2 \tilde{Y}_{\hat{1}\hat{2}} = 0,\ \left(\tilde{e}_0^{\hat{0}}\right)^2 \tilde{Y}_{\hat{2}\hat{3}} = 0,\ \left(\tilde{e}_0^{\hat{0}}\right)^2 \tilde{Y}_{\hat{3}\hat{1}} = 0; \tag{2-57}$$

$$\left(\tilde{e}_0^{\hat{0}}\right)^2 \tilde{Y}_{\hat{1}\hat{1}} - \frac{1}{3}\Phi_1 = 0,\ \left(\tilde{e}_0^{\hat{0}}\right)^2 \tilde{Y}_{\hat{2}\hat{2}} + \frac{1}{2}\left(\frac{1}{3}\Phi_1 - \Phi_2\right) = 0,\ \left(\tilde{e}_0^{\hat{0}}\right)^2 \tilde{Y}_{\hat{3}\hat{3}} + \frac{1}{2}\left(\frac{1}{3}\Phi_1 + \Phi_2\right) = 0; \tag{2-58}$$

According to (2-30) we see that if both $\Phi_1$ and $\Phi_2$ are in the form (2-53), then all six conditions in (2-57) and (2-58) are independent with $\tilde{e}_0^{\hat{0}}$.

In the above nine gauge conditions, the six given by (2-56), (2-57) and the three given by (2-58) are provided by local Lorentz transformation $e_\mu^{\hat{\alpha}} = \Lambda_{\hat{\beta}}^{\hat{\alpha}}(x)\tilde{e}_\mu^{\hat{\beta}}$ and coordinate transformation $x^\mu = x^\mu(\tilde{x}^\nu)$, respectively.

Although coordinate transformation can provide four coordinate conditions, what we have used is only three of four coordinate conditions. As an example, we can add the fourth coordinate condition $g^{00}=-\left(\tilde{e}^{0}_{\hat{0}}\right)^{2}=-\left(\tilde{e}^{\hat{0}}_{0}\right)^{-2}=-1$, since $\tilde{e}^{\hat{0}}_{0}$ is independent with the nine gauge conditions (2-56) ~ (2-58). In fact, we can choose any relation about $g_{\mu\nu}$ and $g_{\rho\sigma,\tau}$ being independent with the three coordinate conditions (2-58), in other words (2-48), as the fourth coordinate condition. However, we would rather to remain one undetermined coordinate condition for other significant purpose, the two functions $\Phi_1$ and $\Phi_2$ as well. Although two chooses have been given in (2-54) and (2-55), we can choose different forms of $\Phi_1$ and $\Phi_2$ for significant purpose.

We now can investigate a system whose basic variables are $\left\{\tilde{e}^{\hat{\gamma}}_{0},\tilde{e}^{\hat{a}}_{i}\right\}$ and the action with matter is

$$S_{(2)}=\int\tilde{e}^{\hat{0}}_{0}(x)\left\|{}^{3}\tilde{e}(x)\right\|\mathrm{d}^{4}xL_{\mathrm{Total}(2)}(x),\quad L_{\mathrm{Total}(2)}=\frac{c^{3}}{16\pi G}\left(L_{\mathrm{GK}(3)}+2\tilde{e}^{0}_{\hat{0}}\tilde{e}^{\hat{0}}_{0,k}U^{k}+L_{\mathrm{GV}}\right)+L_{\mathrm{M}},\quad(2\text{-}59)$$

where $L_{\mathrm{GK}(3)}$ is given by (2-50) but in which $\Phi_1$ and $\Phi_2$ are in the form (2-53), $U^k$ and $L_{\mathrm{GV}}$ are still given by (2-15) but in which $\left\{e^{\hat{a}}_{i}\right\}$ is replaced with $\left\{\tilde{e}^{\hat{a}}_{i}\right\}$.

Using the six gauge conditions (2-57) and (2-58) and according to the proof process of the Theorem in Sect. 2.1, it is easy to prove that the equations of motion of the system obtained by the Euler-Lagrange equations corresponding to the action (2-59) are equivalent to the equations (2-17) ~ (2-24) but in which $S^{\hat{0}i\hat{a}}$ is expressed by (2-49) and in (2-19) and (2-21), the term $\frac{1}{2}L_{\mathrm{GK}}$ is replaced with $\frac{1}{4}\left(\tilde{e}^{0}_{\hat{0}}\right)^{2}\left(\frac{1}{3}\Phi_1^{2}+\Phi_2^{2}\right)$.

For the system described by the action (2-59), the momenta conjugate to $\tilde{e}^{\hat{\gamma}}_{0}$ and $\tilde{e}^{\hat{a}}_{i}$ are

$$\begin{aligned}\tilde{\pi}^{0}_{\hat{\gamma}}&=\frac{\partial\left(\tilde{e}^{\hat{0}}_{0}\left\|{}^{3}\tilde{e}\right\|L_{\mathrm{Total}(2)}\right)}{\partial\,\tilde{e}^{\hat{\gamma}}_{0,0}}=\tilde{e}^{\hat{0}}_{0}\left\|{}^{3}\tilde{e}\right\|\frac{\partial L_{\mathrm{M}}}{\partial\,\tilde{e}^{\hat{\gamma}}_{0,0}},\\ \tilde{\pi}^{i}_{\hat{a}}&=\frac{\partial\left(\tilde{e}^{\hat{0}}_{0}\left\|{}^{3}\tilde{e}\right\|L_{\mathrm{Total}(2)}\right)}{\partial\,\tilde{e}^{\hat{a}}_{i,0}}=\frac{c^{3}}{16\pi G}\tilde{e}^{\hat{0}}_{0}\left\|{}^{3}\tilde{e}\right\|\frac{\partial L_{\mathrm{GK}(3)}}{\partial\,\tilde{e}^{\hat{a}}_{i,0}}+\tilde{e}^{\hat{0}}_{0}\left\|{}^{3}\tilde{e}\right\|\frac{\partial L_{\mathrm{M}}}{\partial\,\tilde{e}^{\hat{a}}_{i,0}},\end{aligned}\qquad(2\text{-}60)$$

by the expression (2-50) of $L_{\mathrm{GK}(3)}$, we obtain the concrete forms of $\tilde{\pi}^{i}_{\hat{a}}$:

$$\begin{aligned}&\tilde{\pi}^{i}_{\hat{1}}=\frac{c^{3}}{24\pi G}\tilde{e}^{0}_{\hat{0}}\left\|{}^{3}\tilde{e}\right\|\tilde{e}^{i}_{\hat{1}}\Phi_1+\tilde{e}^{\hat{0}}_{0}\left\|{}^{3}\tilde{e}\right\|\frac{\partial L_{\mathrm{M}}}{\partial\,\tilde{e}^{\hat{1}}_{i,0}},\quad\tilde{\pi}^{i}_{\hat{2}}=-\frac{c^{3}}{16\pi G}\tilde{e}^{0}_{\hat{0}}\left\|{}^{3}\tilde{e}\right\|\tilde{e}^{i}_{\hat{2}}\left(\frac{1}{3}\Phi_1-\Phi_2\right)+\tilde{e}^{\hat{0}}_{0}\left\|{}^{3}\tilde{e}\right\|\frac{\partial L_{\mathrm{M}}}{\partial\,\tilde{e}^{\hat{2}}_{i,0}},\\ &\tilde{\pi}^{i}_{\hat{3}}=-\frac{c^{3}}{16\pi G}\tilde{e}^{0}_{\hat{0}}\left\|{}^{3}\tilde{e}\right\|\tilde{e}^{i}_{\hat{3}}\left(\frac{1}{3}\Phi_1+\Phi_2\right)+\tilde{e}^{\hat{0}}_{0}\left\|{}^{3}\tilde{e}\right\|\frac{\partial L_{\mathrm{M}}}{\partial\,\tilde{e}^{\hat{3}}_{i,0}}.\end{aligned}\qquad(2\text{-}61)$$

If we want to obtain the corresponding Hamiltonian representation of (2-59), then we first should express $\tilde{e}^{\hat{\gamma}}_{0,0}$ and $\tilde{e}^{\hat{a}}_{i,0}$ via $\tilde{\pi}^{0}_{\hat{\gamma}}$ and $\tilde{\pi}^{i}_{\hat{a}}$, namely, obtain the expressions $\tilde{e}^{\hat{\gamma}}_{0,0}=\tilde{e}^{\hat{\gamma}}_{0,0}\left(\tilde{\pi}^{0}_{\hat{\gamma}},\tilde{\pi}^{i}_{\hat{a}}\right)$ and $\tilde{e}^{\hat{a}}_{i,0}=\tilde{e}^{\hat{a}}_{i,0}\left(\tilde{\pi}^{0}_{\hat{\gamma}},\tilde{\pi}^{i}_{\hat{a}}\right)$ from (2-60) and (2-61). However, generally speaking, there is not any quadratic term of first time derivative $\tilde{e}^{\hat{\gamma}}_{0,0}$ or $\tilde{e}^{\hat{a}}_{i,0}$ in $L_{\mathrm{M}}$ of matter, namely, there is not any time derivative term $\tilde{e}^{\hat{\gamma}}_{0,0}$ or $\tilde{e}^{\hat{a}}_{i,0}$ in $\frac{\partial L_{\mathrm{M}}}{\partial \tilde{e}^{\hat{\gamma}}_{0,0}}$ and $\frac{\partial L_{\mathrm{M}}}{\partial \tilde{e}^{\hat{a}}_{i,0}}$, all the Dirac field, the Klein-Gordon field, the Maxwell field and the Yang-Mills field in curve spacetime have such characteristic. Hence, if both $\Phi_1$ and $\Phi_2$ are in the form (2-54) in which there is not any time derivative term $\tilde{e}^{\hat{\gamma}}_{0,0}$ or $\tilde{e}^{\hat{a}}_{i,0}$, then none of the expressions $\tilde{e}^{\hat{\gamma}}_{0,0} = \tilde{e}^{\hat{\gamma}}_{0,0}\left(\tilde{\pi}^{0}_{\hat{\gamma}}, \tilde{\pi}^{i}_{\hat{a}}\right)$ and $\tilde{e}^{\hat{a}}_{i,0} = \tilde{e}^{\hat{a}}_{i,0}\left(\tilde{\pi}^{0}_{\hat{\gamma}}, \tilde{\pi}^{i}_{\hat{a}}\right)$ can be obtained by (2-60) and (2-61).

As well-known, if the determinant of the matrix $\left[\frac{\partial^2 L(\phi_c, \phi_{c,\tau})}{\partial \dot{\phi}_a \partial \dot{\phi}_b}\right]$ of a Lagrangian density $L(\phi_a, \phi_{a,\lambda})$ vanishes, then the system described by the Lagrangian $L(\phi_a, \phi_{a,\lambda})$ is so called singular Lagrange system. However, for the system whose basic variables and the action are $\left\{\tilde{e}^{\hat{\gamma}}_0, \tilde{e}^{\hat{a}}_i\right\}$ and (2-59), respectively, denoting $\phi_a \equiv \left\{\tilde{e}^{\hat{\gamma}}_0, \tilde{e}^{\hat{a}}_i\right\}$, $L(\phi_a, \phi_{a,\lambda}) \equiv \tilde{e}^{\hat{0}}_0 \left\| {}^3\tilde{e} \right\| L_{\mathrm{Total(2)}}$, where $L_{\mathrm{Total(2)}}$ is given by (2-59), we see that not only the determinant of the matrix $\left[\frac{\partial^2 L(\phi_c, \phi_{c,\tau})}{\partial \dot{\phi}_a \partial \dot{\phi}_b}\right]$ vanishes, but also the rank of $\left[\frac{\partial^2 L(\phi_c, \phi_{c,\tau})}{\partial \dot{\phi}_a \partial \dot{\phi}_b}\right]$ is zero, since from (2-60) and (2-61) we see that every matrix element $\frac{\partial^2 L(\phi_c, \phi_{c,\tau})}{\partial \dot{\phi}_a \partial \dot{\phi}_b} = 0$. We call such system *fully singular Lagrange system.*

Formally, an example of fully singular Lagrange system is the Dirac field in flat spacetime, since from the Lagrange $L_{\mathrm{D}} = \bar{\psi}\left(\mathrm{i}\hbar\gamma^{\mu}\frac{\partial}{\partial x^{\mu}} - mc\right)\psi$ of the Dirac field we have $\pi_{\alpha} = \frac{\partial L_{\mathrm{D}}}{\partial \dot{\psi}_{\alpha}} = \frac{\mathrm{i}\hbar}{c}\psi^{+}_{\alpha}$ and $\frac{\partial^2 L_{\mathrm{D}}}{\partial \dot{\psi}_{\alpha} \partial \dot{\psi}_{\beta}} = 0$.

As well-known, the system described by (2-4) and (2-5) is not equivalent to the theory of electromagnetic field described by (2-3), the equivalence asks to add the Lorenz condition $A^{\lambda}{}_{,\lambda} = 0$, or $\left\langle \middle| A^{\lambda}{}_{,\lambda} \middle| \right\rangle = 0$ for state vector $\left| \ \right\rangle$ in the corresponding quantum theory. Similarly, the fully singular Lagrange system described by (2-59) is not equivalent to general relativity, the equivalence asks to add the additional conditions (2-57) and (2-58), or add the six conditions

$$\left\langle \left| \left(\tilde{e}_0^{\hat{0}}\right)^2 \tilde{Y}_{\hat{1}\hat{2}} \right| \right\rangle = 0\,,\ \left\langle \left| \left(\tilde{e}_0^{\hat{0}}\right)^2 \tilde{Y}_{\hat{2}\hat{3}} \right| \right\rangle = 0\,,\ \left\langle \left| \left(\tilde{e}_0^{\hat{0}}\right)^2 \tilde{Y}_{\hat{3}\hat{1}} \right| \right\rangle = 0\,;\ \left\langle \left| \left( \left(\tilde{e}_0^{\hat{0}}\right)^2 \tilde{Y}_{\hat{1}\hat{1}} - \frac{1}{3}\Phi_1 \right) \right| \right\rangle = 0\,,$$
$$\left\langle \left| \left( \left(\tilde{e}_0^{\hat{0}}\right)^2 \tilde{Y}_{\hat{2}\hat{2}} + \frac{1}{2}\left(\frac{1}{3}\Phi_1 - \Phi_2\right) \right) \right| \right\rangle = 0\,,\ \left\langle \left| \left( \left(\tilde{e}_0^{\hat{0}}\right)^2 \tilde{Y}_{\hat{3}\hat{3}} + \frac{1}{2}\left(\frac{1}{3}\Phi_1 + \Phi_2\right) \right) \right| \right\rangle = 0 \qquad (2\text{-}62)$$

for state vector $\left| \ \right\rangle$ in the corresponding quantum theory, where the expression of $\tilde{Y}_{\hat{a}\hat{b}}$ is given by (2-30).

We can try to use the Dirac-Bargmann method for a singular Lagrangian system or the method of path integral to realize quantization of the fully singular Lagrange system described by (2-59). This will be studied further.

On the other hand, all second time derivative terms in the vierbein forms of the ten Einstein equations can be eliminated, this characteristic shows that general relativity is great different from other fields, e.g., the Yang-Mills field. For non-Abelian gauge field, we can choose appropriate gauge conditions, for example, the space-axial gauge $A_3^a = 0$ [7], to eliminate some dynamic variables. However, no matter how to choose gauge-fixing terms, we cannot remove all second time derivative terms in the equations of motion of the non-Abelian gauge field $A_\mu^a$. Hence, it is impossible to ascribe general relativity to non-Abelian gauge field.

## 3 The generation of the simplest constraint conditions

In this section, we still discuss the Lagrangian $L_{G(2)}$ given by (1-6) and the vierbein forms of the Einstein equations under time gauge condition, hence, all formulas in Sect. 2.2 hold. For the purpose that eliminates all first time derivative terms in $S^{\hat{0}i\hat{a}}$ given by (2-22) and all the quadratic terms of first time derivative in $L_{GK}$ given by (2-9), we are not limited to the simplest constraint conditions discussed in Sect. 2.3 but consider a more general quantity $\Omega_{ij}$:

$$\Omega_{ij} = -\frac{1}{2} g^{00} g_{ij,0} + \overline{\Omega}_{ij}\,,\ \overline{\Omega}_{ij} = \overline{\Omega}_{ij}\left(g_{\mu\nu}\,,\ g_{\mu\nu,k}\right),\ \overline{\Omega}_{ij} = \overline{\Omega}_{ji}\,, \qquad (3\text{-}1)$$

namely, $\overline{\Omega}_{ij}$ is a function of metric tensor $g_{\mu\nu}$ and first space derivative $g_{\mu\nu,k}$, and symmetric in the pair of indices. For example, $\overline{\Omega}_{ij} = 0$, $\overline{\Omega}_{ij} = \frac{1}{2} g^{0k} g_{ij,k}$, etc. And then, we introduce

$$Z_{\hat{a}\hat{b}} = e_{\hat{a}}^i e_{\hat{b}}^j \Omega_{ij} = Y_{\hat{a}\hat{b}} - e_{\hat{a}}^i e_{\hat{b}}^j \omega_{ij}\,, \qquad (3\text{-}2)$$

where $Y_{\hat{a}\hat{b}}$ is given by (2-10); from (2-10) and (2-51) we have

$$\omega_{ij} = \Gamma_{ij}^0 - \Omega_{ij} = -\frac{1}{2} g^{00}\left[\frac{g^{0k}}{g^{00}} g_{ij,k} + \left(\frac{g^{0k}}{g^{00}}\right)_{,i} g_{kj} + \left(\frac{g^{0k}}{g^{00}}\right)_{,j} g_{ki}\right] - \overline{\Omega}_{ij} \equiv \omega_{ij}\left(g_{\mu\nu}\,,\ g_{\mu\nu,k}\right). \qquad (3\text{-}3)$$

If we take $\overline{\Omega}_{ij} = -\frac{1}{2} g^{00}\left[\frac{g^{0k}}{g^{00}} g_{ij,k} + \left(\frac{g^{0k}}{g^{00}}\right)_{,i} g_{kj} + \left(\frac{g^{0k}}{g^{00}}\right)_{,j} g_{ki}\right]$, then $\Omega_{ij} = \Gamma^0_{ij}$, $Z_{\hat{a}\hat{b}} = Y_{\hat{a}\hat{b}}$ and $\omega_{ij} = 0$, we therefore return to the case discussed in Sect. 2.3.

It is obvious that $Z_{\hat{a}\hat{b}}$ is symmetric in the pair of indices and the manner of transformation under the transformation matrix (2-26) is

$$\widetilde{Z}_{\hat{a}\hat{b}} = \widetilde{e}^i_{\hat{a}} \widetilde{e}^j_{\hat{b}} \Omega_{ij} = \Lambda^{\hat{c}}_{\hat{a}} e^i_{\hat{c}} \Lambda^{\hat{d}}_{\hat{b}} e^j_{\hat{d}} \Omega_{ij} = \Lambda^{\hat{c}}_{\hat{a}} \Lambda^{\hat{d}}_{\hat{b}} Z_{\hat{c}\hat{d}} , \tag{3-4}$$

where

$$\widetilde{Z}_{\hat{a}\hat{b}} = \widetilde{Y}_{\hat{a}\hat{b}} - \widetilde{e}^i_{\hat{a}} \widetilde{e}^j_{\hat{b}} \omega_{ij} = \frac{1}{2} \widetilde{e}^{\hat{0}}_0 \widetilde{e}^{\lambda}_{\hat{0}} \left[ \widetilde{e}^i_{\hat{a}} \left( \widetilde{e}_{\hat{b}i,\lambda} - \widetilde{e}_{\hat{b}\lambda,i} \right) + \widetilde{e}^i_{\hat{b}} \left( \widetilde{e}_{\hat{a}i,\lambda} - \widetilde{e}_{\hat{a}\lambda,i} \right) \right] - \widetilde{e}^i_{\hat{a}} \widetilde{e}^j_{\hat{b}} \omega_{ij} , \tag{3-5}$$

in (3-5), we have used the expression (2-30) of $\widetilde{Y}_{\hat{a}\hat{b}}$. Hence, similar to $Y_{\hat{a}\hat{b}}$, for a tetrad $\left\{ e^{\hat{\alpha}}_{\mu} \right\}$ satisfying the time gauge condition, if $Z_{\hat{a}\hat{b}} \neq 0$ $(a, b = 1, 2, 3;\ a \neq b)$, then we can choose $\Lambda^{\hat{b}}_{\hat{a}}$ such that

$$\widetilde{Z}_{\hat{1}\hat{2}} = 0 ,\ \ \widetilde{Z}_{\hat{2}\hat{3}} = 0 ,\ \ \widetilde{Z}_{\hat{3}\hat{1}} = 0 \tag{3-6}$$

for a new group of tetrad $\left\{ \widetilde{e}^{\hat{\alpha}}_{\mu} \right\}$ satisfying the time gauge condition yet.

Substituting $\widetilde{Y}_{\hat{a}\hat{b}} = \widetilde{Z}_{\hat{a}\hat{b}} + \widetilde{e}^i_{\hat{a}} \widetilde{e}^j_{\hat{b}} \omega_{ij}$ obtained by (3-5) to (2-28) and using (3-6) to eliminate the terms $\widetilde{Z}^{\hat{a}\hat{b}}$ $(b \neq a)$ in $S^{\hat{0}i\hat{a}}$, we obtain

$$S^{\hat{0}i\hat{a}} = \widetilde{e}^{\hat{0}}_0 \widetilde{e}^{\hat{a}i} \widetilde{Y}^{\hat{b}}_{\hat{b}} - \widetilde{e}^{\hat{0}}_0 \widetilde{e}^{\hat{a}i} \widetilde{Z}^{\hat{a}\hat{a}} - \widetilde{e}^{\hat{0}}_0 \widetilde{e}^i_{\hat{b}} \widetilde{e}^{\hat{a}l} \widetilde{e}^{\hat{b}m} \omega_{lm} \qquad (a = 1, 2, 3) ; \tag{3-7}$$

On the other hand, substituting $Y_{\hat{a}\hat{b}} = Z_{\hat{a}\hat{b}} + e^i_{\hat{a}} e^j_{\hat{b}} \omega_{ij}$ obtained by (3-5) to (2-9) and using some formulas given in Sect. 2, it is easy to prove that $L_{\mathrm{GK}}$ can be written to the form

$$\begin{aligned} L_{\mathrm{GK}} = & -\frac{2}{3}\left(e^{\hat{0}}_0\right)^2 \left(Y^{\hat{a}}_{\hat{a}}\right)^2 + \left(e^{\hat{0}}_0\right)^2 \left( \frac{1}{6}\left(2Z_{\hat{1}\hat{1}} - Z_{\hat{2}\hat{2}} - Z_{\hat{3}\hat{3}}\right)^2 + \frac{1}{2}\left(Z_{\hat{2}\hat{2}} - Z_{\hat{3}\hat{3}}\right)^2 + 2\left(Z_{\hat{1}\hat{2}}\right)^2 + 2\left(Z_{\hat{2}\hat{3}}\right)^2 + 2\left(Z_{\hat{3}\hat{1}}\right)^2 \right) \\ & + \left(e^{\hat{0}}_0\right)^2 \left( \eta^{\hat{a}\hat{c}} \eta^{\hat{b}\hat{d}} - \frac{1}{3} \eta^{\hat{a}\hat{b}} \eta^{\hat{c}\hat{d}} \right) \left( 2Y_{\hat{a}\hat{b}} - e^i_{\hat{a}} e^j_{\hat{b}} \omega_{ij} \right) e^l_{\hat{c}} e^m_{\hat{d}} \omega_{lm} . \end{aligned} \tag{3-8}$$

If in the above expression, $\left\{ e^{\hat{\gamma}}_0 , e^{\hat{a}}_i \right\}$, $Y_{\hat{a}\hat{b}}$ and $Z_{\hat{a}\hat{b}}$ are replaced with $\left\{ \widetilde{e}^{\hat{\gamma}}_0 , \widetilde{e}^{\hat{a}}_i \right\}$, $\widetilde{Y}_{\hat{a}\hat{b}}$ and $\widetilde{Z}_{\hat{a}\hat{b}}$, respectively, then according to the Theorem proved in Sect. 2.1 and (3-6), we can remove the term $\left(e^{\hat{0}}_0\right)^2 \left( 2\left(\widetilde{Z}_{\hat{1}\hat{2}}\right)^2 + 2\left(\widetilde{Z}_{\hat{2}\hat{3}}\right)^2 + 2\left(\widetilde{Z}_{\hat{3}\hat{1}}\right)^2 \right)$ and, thus, $L_{\mathrm{GK}}$ becomes

$$\begin{aligned} L_{\mathrm{GK}(4)} = & -\frac{2}{3}\left(\widetilde{e}^{\hat{0}}_0\right)^2 \left(\widetilde{Y}^{\hat{a}}_{\hat{a}}\right)^2 + \left(\widetilde{e}^{\hat{0}}_0\right)^2 \left( \frac{1}{6}\left(2\widetilde{Z}_{\hat{1}\hat{1}} - \widetilde{Z}_{\hat{2}\hat{2}} - \widetilde{Z}_{\hat{3}\hat{3}}\right)^2 + \frac{1}{2}\left(\widetilde{Z}_{\hat{2}\hat{2}} - \widetilde{Z}_{\hat{3}\hat{3}}\right)^2 \right) \\ & + \left(\widetilde{e}^{\hat{0}}_0\right)^2 \left( \eta^{\hat{a}\hat{c}} \eta^{\hat{b}\hat{d}} - \frac{1}{3} \eta^{\hat{a}\hat{b}} \eta^{\hat{c}\hat{d}} \right) \left( 2\widetilde{Y}_{\hat{a}\hat{b}} - \widetilde{e}^i_{\hat{a}} \widetilde{e}^j_{\hat{b}} \omega_{ij} \right) \widetilde{e}^l_{\hat{c}} \widetilde{e}^m_{\hat{d}} \omega_{lm} . \end{aligned} \tag{3-9}$$

For the equation

$$\left\| Z_{\hat{a}\hat{b}} - \tau I \right\| = -\tau^3 + d_1\tau^2 - d_2\tau + d_3 = 0 \tag{3-10}$$

of eigenvalues of $3\times3$ symmetric matrix $Z_{\hat{a}\hat{b}}$, the coefficients are

$$\begin{aligned} &d_1 = Z_{\hat{1}\hat{1}} + Z_{\hat{2}\hat{2}} + Z_{\hat{3}\hat{3}} = \eta^{\hat{a}\hat{b}} Z_{\hat{a}\hat{b}} = Z^{\hat{a}}_{\hat{a}} = e^{\hat{a}i} e^{j}_{\hat{a}} \Omega_{ij} = \tilde{g}^{ij}\Omega_{ij}, \\ &d_2 = Z_{\hat{1}\hat{1}}Z_{\hat{2}\hat{2}} + Z_{\hat{1}\hat{1}}Z_{\hat{3}\hat{3}} + Z_{\hat{2}\hat{2}}Z_{\hat{3}\hat{3}} - \left(Z_{\hat{1}\hat{2}}\right)^2 - \left(Z_{\hat{2}\hat{3}}\right)^2 - \left(Z_{\hat{3}\hat{1}}\right)^2 = -\frac{1}{2}\left(\tilde{g}^{il}\tilde{g}^{jm} - \tilde{g}^{ij}\tilde{g}^{lm}\right)\Omega_{ij}\Omega_{lm}, \\ &d_3 = \left\| Z_{\hat{a}\hat{b}} \right\| = \left\| e^{i}_{\hat{a}} e^{j}_{\hat{b}} \Omega_{ij} \right\| = \left\| e^{i}_{\hat{a}} \right\|^2 \left\| \Omega_{ij} \right\| = \left\| \tilde{g}^{lm} \right\| \left\| \Omega_{ij} \right\| = \frac{\left\| \Omega_{ij} \right\|}{\left\| g_{lm} \right\|}, \end{aligned} \tag{3-11}$$

where $\left\| \Omega_{ij} \right\|$ means the determinant of the $3\times3$ matrix $\left[\Omega_{ij}\right]$.

Similar to the three characteristics about $\lambda_{(a)}$ $(a = 1,2,3)$ of the cubic equation (2-37) discussed in Sect. 2.3, all the three roots $\tau_{(a)}$ $(a = 1,2,3)$ of (3-10) are real and functions of $g_{\mu\nu}$, $g_{0i,j}$ and $g_{ij,\tau}$:

$$\tau_{(a)} \equiv \tau_{(a)}(g_{\mu\nu}\ ;\ g_{0i,j}\ , g_{ij,\tau}) \quad (a = 1,2,3)\ ; \tag{3-12}$$

and according to Viète's formulas, we have

$$d_1 = \tau_{(1)} + \tau_{(2)} + \tau_{(3)}\ ,\ d_2 = \tau_{(1)}\tau_{(2)} + \tau_{(1)}\tau_{(3)} + \tau_{(2)}\tau_{(3)}\ ,\ d_3 = \tau_{(1)}\tau_{(2)}\tau_{(3)}\ . \tag{3-13}$$

Similar to $\tilde{Y}_{\hat{a}\hat{a}}$ $(a = 1,2,3)$, for $\tilde{Z}_{\hat{a}\hat{a}}$ $(a = 1,2,3)$ we can prove

$$\tilde{Z}_{\hat{a}\hat{a}} = \tilde{e}^{i}_{\hat{a}}\tilde{e}^{j}_{\hat{a}}\Omega_{ij} = \tau_{(a)} \equiv \tau_{(a)}(g_{\mu\nu}\ ;\ g_{0i,j}\ , g_{ij,\tau}) \quad (a = 1,2,3)\ . \tag{3-14}$$

namely, $\tilde{Z}_{\hat{a}\hat{a}}$ $(a = 1,2,3)$ are just the three roots of the cubic equation (3-10).

So far, all six gauge conditions provided by local Lorentz transformation $e^{\hat{\alpha}}_{\mu} = \Lambda^{\hat{\alpha}}_{\hat{\beta}}(x)\tilde{e}^{\hat{\beta}}_{\mu}$ have run out, from (3-14) we see that all $\tilde{Z}_{\hat{a}\hat{a}}$ $(a = 1,2,3)$ are functions of $g_{\mu\nu}$, $g_{0i,j}$ and $g_{ij,\lambda}$ and independent with second derivative $g_{\mu\nu,\rho,\sigma}$, we therefore can further choose special combinations of $\tilde{Z}_{\hat{a}\hat{a}}$ $(a = 1,2,3)$ as coordinate conditions to eliminate all first time derivative terms in $S^{\hat{0}i\hat{a}}$ expressed by (3-7).

We can employ the coordinate condition (2-42) to eliminate $\tilde{Y}^{\hat{a}}_{\hat{a}}$ in (3-7) and the negative kinetic energy term $-\frac{2}{3}\left(e^{\hat{0}}_{0}\right)^2\left(\tilde{Y}^{\hat{a}}_{\hat{a}}\right)^2$ in (3-9), respectively, however, formally, we consider a more general coordinate condition

$$\widetilde{Y}_{\hat{a}}^{\hat{a}} = \widetilde{Y}_{\hat{1}\hat{1}} + \widetilde{Y}_{\hat{2}\hat{2}} + \widetilde{Y}_{\hat{3}\hat{3}} = \widetilde{g}^{ij}\Gamma_{ij}^{0} = \Xi_0 , \tag{3-15}$$

where $\Xi_0$ is appropriate function of $g_{\mu\nu}$ and first space derivative $g_{\mu\nu,k}$,

$$\Xi_0 \equiv \Xi_0\left(g_{\mu\nu} ;\ g_{\mu\nu,k}\right) . \tag{3-16}$$

Using the coordinate condition (3-15), $S^{\hat{0}i\hat{a}}$ expressed by (3-7) becomes

$$S^{\hat{0}i\hat{a}} = \widetilde{e}_0^{\hat{0}}\widetilde{e}^{\hat{a}i}\Xi_0 - \widetilde{e}_0^{\hat{0}}\widetilde{e}^{\hat{a}i}\widetilde{Z}^{\hat{a}\hat{a}} - \widetilde{e}_0^{\hat{0}}\widetilde{e}_{\hat{b}}^{i}\widetilde{e}^{\hat{a}l}\widetilde{e}^{\hat{b}m}\omega_{lm} ; \tag{3-17}$$

On the other hand, the negative kinetic energy term $-\frac{2}{3}\left(e_0^{\hat{0}}\right)^2\left(\widetilde{Y}_{\hat{a}}^{\hat{a}}\right)^2$ can be written to the form

$$-\frac{2}{3}\left(\widetilde{e}_0^{\hat{0}}\right)^2\left(\widetilde{Y}_{\hat{a}}^{\hat{a}}\right)^2 = -\frac{2}{3}\left(\widetilde{e}_0^{\hat{0}}\right)^2\left(\widetilde{Y}_{\hat{a}}^{\hat{a}} - \Xi_0\right)^2 - \frac{4}{3}\left(\widetilde{e}_0^{\hat{0}}\right)^2 \Xi_0\widetilde{Y}_{\hat{a}}^{\hat{a}} + \frac{2}{3}\left(\widetilde{e}_0^{\hat{0}}\right)^2 \Xi_0^2 ,$$

this term appears in the expression (3-9) of $L_{\mathrm{GK}(4)}$ yet, hence, according to the Theorem proved in Sect. 2.1 and (3-15), we can remove the term $-\frac{2}{3}\left(\widetilde{e}_0^{\hat{0}}\right)^2\left(\widetilde{Y}_{\hat{a}}^{\hat{a}} - \Xi_0\right)^2$ in the expression (3-9) and, thus, $L_{\mathrm{GK}(4)}$ becomes

$$\begin{aligned} L_{\mathrm{GK}(5)} = & -\frac{4}{3}\left(\widetilde{e}_0^{\hat{0}}\right)^2 \Xi_0\widetilde{Y}_{\hat{a}}^{\hat{a}} + \frac{2}{3}\left(\widetilde{e}_0^{\hat{0}}\right)^2 \Xi_0^2 + \left(\widetilde{e}_0^{\hat{0}}\right)^2\left(\frac{1}{6}\left(2\widetilde{Z}_{\hat{1}\hat{1}} - \widetilde{Z}_{\hat{2}\hat{2}} - \widetilde{Z}_{\hat{3}\hat{3}}\right)^2 + \frac{1}{2}\left(\widetilde{Z}_{\hat{2}\hat{2}} - \widetilde{Z}_{\hat{3}\hat{3}}\right)^2\right) \\ & + \left(\widetilde{e}_0^{\hat{0}}\right)^2\left(\eta^{\hat{a}\hat{c}}\eta^{\hat{b}\hat{d}} - \frac{1}{3}\eta^{\hat{a}\hat{b}}\eta^{\hat{c}\hat{d}}\right)\left(2\widetilde{Y}_{\hat{a}\hat{b}} - \widetilde{e}_{\hat{a}}^{i}\widetilde{e}_{\hat{b}}^{j}\omega_{ij}\right)\widetilde{e}_{\hat{c}}^{l}\widetilde{e}_{\hat{d}}^{m}\omega_{lm} . \end{aligned} \tag{3-18}$$

According to (3-5) and considering $\widetilde{g}^{ij} = \widetilde{e}_{\hat{a}}^{i}\widetilde{e}^{\hat{a}j}$, the coordinate condition (3-15) can be written to the form

$$\widetilde{Z}_{\hat{1}\hat{1}} + \widetilde{Z}_{\hat{2}\hat{2}} + \widetilde{Z}_{\hat{3}\hat{3}} = \Xi_0 - \widetilde{g}^{ij}\omega_{ij} . \tag{3-19}$$

We further choose the following two coordinate conditions

$$2\widetilde{Z}_{\hat{1}\hat{1}} - \widetilde{Z}_{\hat{2}\hat{2}} - \widetilde{Z}_{\hat{3}\hat{3}} = \Xi_1 , \tag{3-20}$$

$$\widetilde{Z}_{\hat{2}\hat{2}} - \widetilde{Z}_{\hat{3}\hat{3}} = \Xi_2 , \tag{3-21}$$

where $\Xi_1$ and $\Xi_2$ are appropriate functions of $g_{\mu\nu}$ and first space derivative $g_{\mu\nu,k}$,

$$\Xi_1 \equiv \Xi_1\left(g_{\mu\nu} ;\ g_{\mu\nu,k}\right) , \quad \Xi_2 \equiv \Xi_2\left(g_{\mu\nu} ;\ g_{\mu\nu,k}\right) . \tag{3-22}$$

Considering (3-14), the three coordinate conditions (3-19) ~ (3-21) can be written to the following forms:

$$\tilde{Z}_{\hat{1}\hat{1}} = \tau_{(1)} = -\frac{1}{3}\tilde{g}^{ij}\omega_{ij} + \frac{1}{3}(\Xi_0 + \Xi_1),\ \tilde{Z}_{\hat{2}\hat{2}} = \tau_{(2)} = -\frac{1}{3}\tilde{g}^{ij}\omega_{ij} + \frac{1}{2}\left(\frac{2}{3}\Xi_0 - \frac{1}{3}\Xi_1 + \Xi_2\right),$$
$$\tilde{Z}_{\hat{3}\hat{3}} = \tau_{(3)} = -\frac{1}{3}\tilde{g}^{ij}\omega_{ij} + \frac{1}{2}\left(\frac{2}{3}\Xi_0 - \frac{1}{3}\Xi_1 - \Xi_2\right);\qquad (3\text{-}23)$$

According to (3-11) and (3-13), the above three coordinate conditions are equivalent to

$$\tilde{g}^{ij}\Omega_{ij} = -\tilde{g}^{ij}\omega_{ij} + \Xi_0,$$
$$\left(\tilde{g}^{il}\tilde{g}^{jm} - \tilde{g}^{ij}\tilde{g}^{lm}\right)\Omega_{ij}\Omega_{lm} = -\frac{2}{3}\left(\tilde{g}^{ij}\omega_{ij} - \Xi_0\right)^2 + \frac{1}{2}\left(\frac{1}{3}\Xi_1^2 + \Xi_2^2\right),\qquad (3\text{-}24)$$
$$\frac{\|\Omega_{ij}\|}{\|g_{lm}\|} = \frac{1}{3}\left(-\tilde{g}^{ij}\omega_{ij} + \Xi_0 + \Xi_1\right)\left(\frac{1}{9}\left(\tilde{g}^{ij}\omega_{ij} - \Xi_0 + \frac{1}{2}\Xi_1\right)^2 - \frac{1}{4}\Xi_2^2\right);$$

Notice that the first condition in (3-24) is just $\tilde{g}^{ij}\Gamma_{ij}^0 = \Xi_0$, and substituting $\Omega_{ij} = \Gamma_{ij}^0 - \omega_{ij}$ obtained by (3-3) to the above three coordinate conditions, (3-24) thus be written to the form:

$$\tilde{g}^{ij}\Gamma_{ij}^0 = \Xi_0,\ \tilde{g}^{il}\tilde{g}^{jm}\left(\Gamma_{ij}^0 - \omega_{ij}\right)\left(\Gamma_{lm}^0 - \omega_{lm}\right) - \frac{1}{3}\left(\tilde{g}^{ij}\omega_{ij} - \Xi_0\right)^2 = \frac{1}{2}\left(\frac{1}{3}\Xi_1^2 + \Xi_2^2\right),$$
$$\frac{\|\Gamma_{ij}^0 - \omega_{ij}\|}{\|g_{lm}\|} = \frac{1}{3}\left(-\tilde{g}^{ij}\omega_{ij} + \Xi_0 + \Xi_1\right)\left(\frac{1}{9}\left(\tilde{g}^{ij}\omega_{ij} - \Xi_0 + \frac{1}{2}\Xi_1\right)^2 - \frac{1}{4}\Xi_2^2\right).\qquad (3\text{-}25)$$

Under the three coordinate conditions (3-23) and considering $\tilde{g}^{lm} = \tilde{e}_{\hat{a}}^l\tilde{e}^{\hat{a}m}$ in (2-8), $S^{\hat{0}i\hat{a}}$ given by (3-17) becomes

$$S^{\hat{0}i\hat{1}} = \tilde{e}_0^{\hat{0}}\tilde{e}^{\hat{1}i}\Xi_0 - \tilde{e}_0^{\hat{0}}\tilde{e}^{\hat{1}i}\left(-\frac{1}{3}\tilde{g}^{lm}\omega_{lm} + \frac{1}{3}\Xi_0 + \frac{1}{3}\Xi_1\right) - \tilde{e}_0^{\hat{0}}\tilde{e}_{\hat{b}}^i\tilde{e}^{\hat{1}l}\tilde{e}^{\hat{b}m}\omega_{lm}$$
$$= \frac{1}{3}\tilde{e}_0^{\hat{0}}\tilde{e}^{\hat{1}i}(2\Xi_0 - \Xi_1) + \tilde{e}_0^{\hat{0}}\left(\frac{1}{3}\tilde{e}^{\hat{1}i}\tilde{e}_{\hat{a}}^l - \tilde{e}_{\hat{a}}^i\tilde{e}^{\hat{1}l}\right)\tilde{e}^{\hat{a}m}\omega_{lm},$$
$$S^{\hat{0}i\hat{2}} = \tilde{e}_0^{\hat{0}}\tilde{e}^{\hat{2}i}\Xi_0 - \tilde{e}_0^{\hat{0}}\tilde{e}^{\hat{2}i}\left(-\frac{1}{3}\tilde{g}^{ij}\omega_{ij} + \frac{1}{2}\left(\frac{2}{3}\Xi_0 - \frac{1}{3}\Xi_1 + \Xi_2\right)\right) - \tilde{e}_0^{\hat{0}}\tilde{e}_{\hat{b}}^i\tilde{e}^{\hat{2}l}\tilde{e}^{\hat{b}m}\omega_{lm}$$
$$= \tilde{e}_0^{\hat{0}}\tilde{e}^{\hat{2}i}\left(\frac{2}{3}\Xi_0 + \frac{1}{6}\Xi_1 - \frac{1}{2}\Xi_2\right) + \tilde{e}_0^{\hat{0}}\left(\frac{1}{3}\tilde{e}^{\hat{2}i}\tilde{e}_{\hat{a}}^l - \tilde{e}_{\hat{a}}^i\tilde{e}^{\hat{2}l}\right)\tilde{e}^{\hat{a}m}\omega_{lm},\qquad (3\text{-}26)$$
$$S^{\hat{0}i\hat{3}} = \tilde{e}_0^{\hat{0}}\tilde{e}^{\hat{3}i}\Xi_0 - \tilde{e}_0^{\hat{0}}\tilde{e}^{\hat{3}i}\left(-\frac{1}{3}\tilde{g}^{ij}\omega_{ij} + \frac{1}{2}\left(\frac{2}{3}\Xi_0 - \frac{1}{3}\Xi_1 - \Xi_2\right)\right) - \tilde{e}_0^{\hat{0}}\tilde{e}_{\hat{b}}^i\tilde{e}^{\hat{3}l}\tilde{e}^{\hat{b}m}\omega_{lm}$$
$$= \tilde{e}_0^{\hat{0}}\tilde{e}^{\hat{3}i}\left(\frac{2}{3}\Xi_0 + \frac{1}{6}\Xi_1 + \frac{1}{2}\Xi_2\right) + \tilde{e}_0^{\hat{0}}\left(\frac{1}{3}\tilde{e}^{\hat{3}i}\tilde{e}_{\hat{a}}^l - \tilde{e}_{\hat{a}}^i\tilde{e}^{\hat{3}l}\right)\tilde{e}^{\hat{a}m}\omega_{lm}.$$

In Sect. 2.5 we point out that not both $\Phi_1$ and $\Phi_2$ introduced by (2-45) and (2-46) vanish, however, even if we now take $\Xi_0 = \Xi_1 = \Xi_2 = 0$, it is not incompatible with the equation $e_0^{\hat{0}}\Theta_{\hat{a}}^0 = 0$ since from (3-26) we see that $S^{\hat{0}i\hat{a}} \neq 0$ when $\Xi_0 = \Xi_1 = \Xi_2 = 0$ but $\omega_{ij} \neq 0$.

On the other hand, we rewrite $L_{\mathrm{GK}(5)}$ given by (3-15) to the form

$$L_{\mathrm{GK}(5)}=-\frac{4}{3}\left(\tilde{e}_0^{\hat{0}}\right)^2 \Xi_0\tilde{Y}_{\hat{a}}^{\hat{a}}+\frac{2}{3}\left(\tilde{e}_0^{\hat{0}}\right)^2\Xi_0^2+\frac{1}{6}\left(\tilde{e}_0^{\hat{0}}\right)^2\left(2\tilde{Z}_{\hat{1}\hat{1}}-\tilde{Z}_{\hat{2}\hat{2}}-\tilde{Z}_{\hat{3}\hat{3}}-\Xi_1\right)^2+\frac{1}{2}\left(\tilde{e}_0^{\hat{0}}\right)^2\left(\tilde{Z}_{\hat{2}\hat{2}}-\tilde{Z}_{\hat{3}\hat{3}}-\Xi_2\right)^2$$
$$+\left(\tilde{e}_0^{\hat{0}}\right)^2\left(\frac{1}{3}\left(2\tilde{Z}_{\hat{1}\hat{1}}-\tilde{Z}_{\hat{2}\hat{2}}-\tilde{Z}_{\hat{3}\hat{3}}\right)\Xi_1+\left(\tilde{Z}_{\hat{2}\hat{2}}-\tilde{Z}_{\hat{3}\hat{3}}\right)\Xi_2\right)-\frac{1}{2}\left(\tilde{e}_0^{\hat{0}}\right)^2\left(\frac{1}{3}\Xi_1^2+\Xi_2^2\right)$$
$$+\left(\tilde{e}_0^{\hat{0}}\right)^2\left(\eta^{\hat{a}\hat{c}}\eta^{\hat{b}\hat{d}}-\frac{1}{3}\eta^{\hat{a}\hat{b}}\eta^{\hat{c}\hat{d}}\right)\left(2\tilde{Y}_{\hat{a}\hat{b}}-\tilde{e}_{\hat{a}}^i\tilde{e}_{\hat{b}}^j\omega_{ij}\right)\tilde{e}_{\hat{c}}^l\tilde{e}_{\hat{d}}^m\omega_{lm}\,,$$

according to the Theorem proved in Sect. 2.1 and (3-20) and (3-21), in the above expression both the terms $\frac{1}{6}\left(\tilde{e}_0^{\hat{0}}\right)^2\left(2\tilde{Z}_{\hat{1}\hat{1}}-\tilde{Z}_{\hat{2}\hat{2}}-\tilde{Z}_{\hat{3}\hat{3}}-\Xi_1\right)^2$ and $\frac{1}{2}\left(\tilde{e}_0^{\hat{0}}\right)^2\left(\tilde{Z}_{\hat{2}\hat{2}}-\tilde{Z}_{\hat{3}\hat{3}}-\Xi_2\right)^2$ can be removed directly, and, further, using (3-5), $L_{\mathrm{GK}(5)}$ becomes

$$L_{\mathrm{GK}(6)}=-\frac{4}{3}\left(\tilde{e}_0^{\hat{0}}\right)^2\Xi_0\tilde{Y}_{\hat{a}}^{\hat{a}}+\frac{1}{3}\left(\tilde{e}_0^{\hat{0}}\right)^2\left(2\tilde{Y}_{\hat{1}\hat{1}}-\tilde{Y}_{\hat{2}\hat{2}}-\tilde{Y}_{\hat{3}\hat{3}}\right)\left(\Xi_1+2\tilde{e}_{\hat{1}}^i\tilde{e}_{\hat{1}}^j\omega_{ij}-\tilde{e}_{\hat{2}}^i\tilde{e}_{\hat{2}}^j\omega_{ij}-\tilde{e}_{\hat{3}}^i\tilde{e}_{\hat{3}}^j\omega_{ij}\right)$$
$$+\left(\tilde{e}_0^{\hat{0}}\right)^2\left(\tilde{Y}_{\hat{2}\hat{2}}-\tilde{Y}_{\hat{3}\hat{3}}\right)\left(\Xi_2+\tilde{e}_{\hat{2}}^i\tilde{e}_{\hat{2}}^j\omega_{ij}-\tilde{e}_{\hat{3}}^i\tilde{e}_{\hat{3}}^j\omega_{ij}\right)+4\left(\tilde{e}_0^{\hat{0}}\right)^2\left(\tilde{Y}_{\hat{1}\hat{2}}\tilde{e}_{\hat{1}}^i\tilde{e}_{\hat{2}}^j\omega_{ij}+\tilde{Y}_{\hat{2}\hat{3}}\tilde{e}_{\hat{2}}^i\tilde{e}_{\hat{3}}^j\omega_{ij}+\tilde{Y}_{\hat{3}\hat{1}}\tilde{e}_{\hat{3}}^i\tilde{e}_{\hat{1}}^j\omega_{ij}\right)$$
$$+\frac{2}{3}\left(\tilde{e}_0^{\hat{0}}\right)^2\Xi_0^2-\frac{1}{6}\left(\tilde{e}_0^{\hat{0}}\right)^2\left(\Xi_1+2\tilde{e}_{\hat{1}}^i\tilde{e}_{\hat{1}}^j\omega_{ij}-\tilde{e}_{\hat{2}}^i\tilde{e}_{\hat{2}}^j\omega_{ij}-\tilde{e}_{\hat{3}}^i\tilde{e}_{\hat{3}}^j\omega_{ij}\right)^2$$
$$-\frac{1}{2}\left(\tilde{e}_0^{\hat{0}}\right)^2\left(\Xi_2+\tilde{e}_{\hat{2}}^i\tilde{e}_{\hat{2}}^j\omega_{ij}-\tilde{e}_{\hat{3}}^i\tilde{e}_{\hat{3}}^j\omega_{ij}\right)^2-2\left(\tilde{e}_0^{\hat{0}}\right)^2\left(\left(\tilde{e}_{\hat{1}}^i\tilde{e}_{\hat{2}}^j\omega_{ij}\right)^2+\left(\tilde{e}_{\hat{2}}^i\tilde{e}_{\hat{3}}^j\omega_{ij}\right)^2+\left(\tilde{e}_{\hat{3}}^i\tilde{e}_{\hat{1}}^j\omega_{ij}\right)^2\right).\quad(3\text{-}27)$$

If $L_{\mathrm{GK}(6)}$ appears in equation of motion, for example, the term $\frac{1}{2}L_{\mathrm{GK}(6)}$ in the equations $e_0^{\hat{0}}\Theta_{\hat{0}}^0=0$ and $e_0^{\hat{0}}\Theta_{\hat{a}}^i=0$ given by (2-17) ~ (2-24), then we can substitute (3-5), (3-6) and (3-23) to (3-27) immediately and, thus, the term $\frac{1}{2}L_{\mathrm{GK}(6)}$ in (2-19) and (2-21) becomes

$$-\frac{1}{3}\left(\tilde{e}_0^{\hat{0}}\right)^2\Xi_0^2+\frac{1}{12}\left(\tilde{e}_0^{\hat{0}}\right)^2\left(\Xi_1+2\tilde{e}_{\hat{1}}^i\tilde{e}_{\hat{1}}^j\omega_{ij}-\tilde{e}_{\hat{2}}^i\tilde{e}_{\hat{2}}^j\omega_{ij}-\tilde{e}_{\hat{3}}^i\tilde{e}_{\hat{3}}^j\omega_{ij}\right)^2$$
$$+\frac{1}{4}\left(\tilde{e}_0^{\hat{0}}\right)^2\left(\Xi_2+\tilde{e}_{\hat{2}}^i\tilde{e}_{\hat{2}}^j\omega_{ij}-\tilde{e}_{\hat{3}}^i\tilde{e}_{\hat{3}}^j\omega_{ij}\right)^2+\left(\tilde{e}_0^{\hat{0}}\right)^2\left(\left(\tilde{e}_{\hat{1}}^i\tilde{e}_{\hat{2}}^j\omega_{ij}\right)^2+\left(\tilde{e}_{\hat{2}}^i\tilde{e}_{\hat{3}}^j\omega_{ij}\right)^2+\left(\tilde{e}_{\hat{3}}^i\tilde{e}_{\hat{1}}^j\omega_{ij}\right)^2\right).\quad(3\text{-}28)$$

Considering $g_{00}=\tilde{e}_0^{\hat{\gamma}}\tilde{e}_{\hat{\gamma}0}$, $g_{0i}=\tilde{e}_0^{\hat{a}}\tilde{e}_{\hat{a}i}$ and $g_{ij}=\tilde{e}_i^{\hat{a}}\tilde{e}_{\hat{a}j}$, $\omega_{ij}$, $\Xi_0$, $\Xi_1$ and $\Xi_2$ introduced by (3-3), (3-16) and (3-22) become

$$\omega_{ij}\equiv\omega_{ij}\left(\tilde{e}_0^{\hat{\gamma}},\ \tilde{e}_k^{\hat{a}}\ ;\ \tilde{e}_{0,n}^{\hat{\gamma}},\ \tilde{e}_{l,m}^{\hat{a}}\right),\ \ \Xi_0\equiv\Xi_0\left(\tilde{e}_0^{\hat{\gamma}},\ \tilde{e}_k^{\hat{a}}\ ;\ \tilde{e}_{0,n}^{\hat{\gamma}},\ \tilde{e}_{l,m}^{\hat{a}}\right),$$
$$\Xi_1\equiv\Xi_1\left(\tilde{e}_0^{\hat{\gamma}},\ \tilde{e}_k^{\hat{a}}\ ;\ \tilde{e}_{0,n}^{\hat{\gamma}},\ \tilde{e}_{l,m}^{\hat{a}}\right),\ \ \Xi_2\equiv\Xi_2\left(\tilde{e}_0^{\hat{\gamma}},\ \tilde{e}_k^{\hat{a}}\ ;\ \tilde{e}_{0,n}^{\hat{\gamma}},\ \tilde{e}_{l,m}^{\hat{a}}\right),\quad(3\text{-}29)$$

Substituting (3-29) to (3-26) and (3-27), we see that there is not any time derivative term $\tilde{e}_{0,0}^{\hat{\gamma}}$ or $\tilde{e}_{i,0}^{\hat{a}}$ in $S^{\hat{0}i\hat{a}}$ and there is not any quadratic term of first time derivative in $L_{\mathrm{GK}(6)}$.

We now can investigate a system whose basic variables are $\left\{\tilde{e}_0^{\hat{\gamma}},\tilde{e}_i^{\hat{a}}\right\}$ and the action with matter is

$$S_{(3)}=\int\tilde{e}_0^{\hat{0}}(x)\left\|{}^3\tilde{e}(x)\right\|\mathrm{d}^4xL_{\mathrm{Total}(3)}(x)\,,\quad L_{\mathrm{Total}(3)}=\frac{c^3}{16\pi G}\left(L_{\mathrm{GK}(6)}+2\tilde{e}_0^0\tilde{e}_{0,k}^{\hat{0}}U^k+L_{\mathrm{GV}}\right)+L_{\mathrm{M}}\,,\quad(3\text{-}30)$$

where $L_{\mathrm{GK}(6)}$ is given by (3-27), $U^k$ and $L_{\mathrm{GV}}$ are still given by (2-15) but in which $\left\{e_i^{\hat{a}}\right\}$ is replaced with $\left\{\tilde{e}_i^{\hat{a}}\right\}$.

From (3-30) we obtain the momenta conjugate to $\tilde{e}_0^{\hat{\gamma}}$ and $\tilde{e}_i^{\hat{a}}$ are

$$
\begin{aligned}
\tilde{\pi}_{\hat{\gamma}}^0 &= \frac{\partial\left(\tilde{e}_0^{\hat{0}}\left\|{}^3\tilde{e}\right\| L_{\mathrm{Total}(3)}\right)}{\partial \tilde{e}_{0,0}^{\hat{\gamma}}} = \tilde{e}_0^{\hat{0}}\left\|{}^3\tilde{e}\right\| \frac{\partial L_{\mathrm{M}}}{\partial \tilde{e}_{0,0}^{\hat{\gamma}}} , \\
\tilde{\pi}_{\hat{1}}^i &= \frac{\partial\left(\tilde{e}_0^{\hat{0}}\left\|{}^3\tilde{e}\right\| L_{\mathrm{Total}(3)}\right)}{\partial \tilde{e}_{i,0}^{\hat{1}}} = \tilde{e}_0^{\hat{0}}\left\|{}^3\tilde{e}\right\| \left(\frac{c^3}{8\pi G}\left(\frac{1}{3}\tilde{e}_{\hat{1}}^i\left(-2\Xi_0+\Xi_1-\tilde{g}^{lm}\omega_{lm}\right)+\tilde{g}^{ij}\tilde{e}_{\hat{1}}^k\omega_{jk}\right)+\frac{\partial L_{\mathrm{M}}}{\partial \tilde{e}_{i,0}^{\hat{1}}}\right) , \\
\tilde{\pi}_{\hat{2}}^i &= \frac{\partial\left(\tilde{e}_0^{\hat{0}}\left\|{}^3\tilde{e}\right\| L_{\mathrm{Total}(3)}\right)}{\partial \tilde{e}_{i,0}^{\hat{2}}} = \tilde{e}_0^{\hat{0}}\left\|{}^3\tilde{e}\right\| \left(\frac{c^3}{8\pi G}\left(\frac{1}{3}\tilde{e}_{\hat{2}}^i\left(-2\Xi_0-\frac{1}{2}\Xi_1+\frac{3}{2}\Xi_2-\tilde{g}^{lm}\omega_{lm}\right)+\tilde{g}^{ij}\tilde{e}_{\hat{2}}^k\omega_{jk}\right)+\frac{\partial L_{\mathrm{M}}}{\partial \tilde{e}_{i,0}^{\hat{2}}}\right) , \\
\tilde{\pi}_{\hat{3}}^i &= \frac{\partial\left(\tilde{e}_0^{\hat{0}}\left\|{}^3\tilde{e}\right\| L_{\mathrm{Total}(3)}\right)}{\partial \tilde{e}_{i,0}^{\hat{3}}} = \tilde{e}_0^{\hat{0}}\left\|{}^3\tilde{e}\right\| \left(\frac{c^3}{8\pi G}\left(\frac{1}{3}\tilde{e}_{\hat{3}}^i\left(-2\Xi_0-\frac{1}{2}\Xi_1-\frac{3}{2}\Xi_2-\tilde{g}^{lm}\omega_{lm}\right)+\tilde{g}^{ij}\tilde{e}_{\hat{3}}^k\omega_{jk}\right)+\frac{\partial L_{\mathrm{M}}}{\partial \tilde{e}_{i,0}^{\hat{3}}}\right) .
\end{aligned}
\tag{3-31}
$$

Hence, if $\omega_{ij}$, $\Xi_0$, $\Xi_1$ and $\Xi_2$ are expressed by (3-29) in which there is not any time derivative term $\tilde{e}_{0,0}^{\hat{\gamma}}$ or $\tilde{e}_{i,0}^{\hat{a}}$ and there is not any quadratic term of first time derivative $\tilde{e}_{0,0}^{\hat{\gamma}}$ or $\tilde{e}_{i,0}^{\hat{a}}$ in $L_{\mathrm{M}}$ of matter, then none of the expressions $\tilde{e}_{0,0}^{\hat{\gamma}}=\tilde{e}_{0,0}^{\hat{\gamma}}\left(\tilde{\pi}_{\hat{\gamma}}^0,\tilde{\pi}_{\hat{a}}^i\right)$ and $\tilde{e}_{i,0}^{\hat{a}}=\tilde{e}_{i,0}^{\hat{a}}\left(\tilde{\pi}_{\hat{\gamma}}^0,\tilde{\pi}_{\hat{a}}^i\right)$ can be obtained by (3-31). And, further, similar to the system described by the action (2-59), the system described by the action (3-30) is also a fully singular Lagrange system and equivalent to general relativity under the six conditions (3-6) and (3-23) but in which all $\omega_{ij}$, $\Xi_0$, $\Xi_1$ and $\Xi_2$ are expressed by (3-29), or six conditions

$$
\left\langle\left|\tilde{Z}_{\hat{1}\hat{2}}\right|\right\rangle=0,\ \left\langle\left|\tilde{Z}_{\hat{2}\hat{3}}\right|\right\rangle=0,\ \left\langle\left|\tilde{Z}_{\hat{3}\hat{1}}\right|\right\rangle=0;\ \left\langle\left|\left(\tilde{Z}_{\hat{1}\hat{1}}+\frac{1}{3}\tilde{g}^{ij}\omega_{ij}-\frac{1}{3}\left(\Xi_0+\Xi_1\right)\right)\right|\right\rangle=0,
$$

$$
\left\langle\left|\left(\tilde{Z}_{\hat{2}\hat{2}}+\frac{1}{3}\tilde{g}^{ij}\omega_{ij}-\frac{1}{2}\left(\frac{2}{3}\Xi_0-\frac{1}{3}\Xi_1+\Xi_2\right)\right)\right|\right\rangle=0,\ \left\langle\left|\left(\tilde{Z}_{\hat{3}\hat{3}}+\frac{1}{3}\tilde{g}^{ij}\omega_{ij}-\frac{1}{2}\left(\frac{2}{3}\Xi_0-\frac{1}{3}\Xi_1-\Xi_2\right)\right)\right|\right\rangle=0
$$

for state vector $|\ \rangle$ in the corresponding quantum theory, where the expression of $\tilde{Z}_{\hat{a}\hat{b}}$ is given by (3-5). For example, under the six gauge conditions (3-6) and (3-23) and according to the proof process of the Theorem in Sect. 2.1, it is easy to prove that the equations of motion of the system obtained by the Euler-Lagrange equations corresponding to the action (3-30) are equivalent to the equations (2-17) ~ (2-24) but in which $S^{\hat{0}i\hat{a}}$ is expressed by (3-26) and in (2-19) and (2-21), the term $\frac{1}{2}L_{\mathrm{GK}}$ is replaced with (3-28).

We can try to choose appropriate functions $\omega_{ij}$, $\Xi_0$, $\Xi_1$ and $\Xi_2$ to simplify the system described by the action (3-30), this will be studied further.